\documentclass[lineno]{jfm}

\usepackage{graphicx}
\usepackage{newtxtext}
\usepackage{newtxmath}
\usepackage{natbib}
\usepackage{hyperref}
\usepackage[dvipsnames]{xcolor}
\hypersetup{
    colorlinks = true,
    urlcolor   = blue,
    citecolor  = MidnightBlue,
}

\newcommand{\RomanNumeralCaps}[1]
\linenumbers

\usepackage{hvfloat}
\usepackage{newfloat}
\DeclareFloatingEnvironment[name=Example,placement=htbp,fileext=lop]{example}
\usepackage{caption}
\newsavebox{\myimage}

\usepackage{soul}

\usepackage{svg}

\usepackage{comment}

\usepackage{csquotes}

\shorttitle{Standing-Wave Dynamics in Low-Frequency Breathing of a TSB}
\shortauthor{L.M. Fuchs, B. Steinfurth, J.G.R.~von Saldern, J. Weiss and K. Oberleithner}

\title{Standing-Wave Dynamics in Low-Frequency Breathing of a Turbulent Separation Bubble}

\author{Lukas M. Fuchs\aff{1}\corresp{\email{l.fuchs@tu-berlin.de}}, B. Steinfurth\aff{2}, J.G.R.~von Saldern\aff{1}, J. Weiss\aff{2} \and K. Oberleithner\aff{1}}

\affiliation{\aff{1}Laboratory for Flow Instabilities and Dynamics, Institute of Fluid Dynamics and Technical Acoustics, Technical University Berlin, Germany
\aff{2}Aerodynamics Laboratory, Technical University Berlin, Germany}

\begin{document}

\newcommand{\ZY}[1]{\textcolor{blue}{#1}}

\maketitle

\begin{abstract}

This study investigates the low-frequency dynamics of a turbulent separation bubble (TSB) forming over a backward-facing ramp, with a focus on large-scale coherent structures associated with the so-called 'breathing motion'. Using time-resolved particle image velocimetry (PIV) in both streamwise and spanwise planes, we examine the role of sidewall confinement, an aspect largely overlooked in previous research. Spectral proper orthogonal decomposition (SPOD) of the streamwise velocity field reveals a dominant low-rank mode at low Strouhal numbers ($St < 0.05$), consistent with prior observations of TSB breathing. Strikingly, the spanwise-oriented PIV data uncover a previously unreported standing wave pattern, characterised by discrete spanwise wavenumbers and nodal/antinodal structures, suggesting the presence of spanwise resonance. To explain these observations, we construct a resolvent-based model that imposes free-slip conditions at the sidewall locations by superposing left- and right-traveling three-dimensional modes. The model accurately reproduces the spanwise structure and frequency content of the measured SPOD modes, demonstrating that sidewall reflections lead to the formation of standing wave-like patterns. Global stability analysis reveals a zero-frequency eigenmode originating from a centrifugal instability, giving rise to the observed low-frequency breathing. Downstream, the associated coherent structures are further amplified through non-modal lift-up mechanisms. Our findings highlight the critical influence of spanwise boundary conditions on the selection and structure of low-frequency modes in TSBs. This has direct implications for both experimental and numerical studies relying on spanwise-periodic boundary conditions and offers a low-order framework for predicting sidewall-induced modal dynamics in separated flows.

\end{abstract}

\begin{keywords}
will be added during the typesetting process
\end{keywords}

\section{Introduction}
When a turbulent boundary layer separates from a surface and reattaches further downstream, the resulting recirculation region is known as a turbulent separation bubble (TSB). These flows exhibit strong unsteadiness over a broad range of frequencies, leading to detrimental effects such as structural vibrations, noise generation, mechanical fatigue, and fluctuations in thermal loads, all of which pose significant challenges in various engineering applications.

TSBs can be classified into two main categories: geometry-induced and adverse pressure gradient (APG)-induced separation bubbles. Geometry-induced TSBs arise at sharp geometric discontinuities such as backward-facing steps \citep{EatonJohnston1982} and rectangular leading-edge configurations \citep{CherryHillier_1984_JFM, Kiya_Sasaki_1983}. APG-induced TSBs occur in configurations such as airfoils \citep{Delery1985_shock}, backward-facing ramps \citep{Kaltenbach1999StudySimulation, WeissAIAAJ2022}, or flat plates where an adverse pressure gradient drives separation \citep{Patrick1987_Nasa, NaMoin1998JFM, Wu2019, Cura2024, Cura2025, Abe2017, MohammedTaifur_Weiss_2016}. One specific instance of an APG-induced TSB is the trailing-edge separation observed on airfoils, where the flow detaches in the rear part of the airfoil and reattaches near the trailing edge. Low-frequency oscillations of these trailing-edge TSBs were first noted by \citet{ZamanMcKinizieRumsey1989JfM} and have since remained a topic of ongoing research \citep{WandGhaemi2022_JfM, SarrasJfM2024}.

Another flow configuration related to TSBs involves stall cells, three-dimensional flow structures that appear during separation on two-dimensional airfoils. These structures have been described as mushroom-shaped \citep{WinkelmannBarlow1980_stallCells}, consisting of two counter-rotating vortices. Additionally, TSBs are prevalent in shock–boundary layer interactions (SBLI), where shock waves interact with turbulent boundary layers, leading to separation and reattachment phenomena that remain a subject of ongoing research \citep{Delery1985_shock, DussaugeDupontDebieve2006, PoggieAIAA2015, Hao2023,Bugeat2022JfM}.

Turbulent separation bubbles have been studied extensively over the past five decades across a wide range of geometries and flow configurations. Throughout these studies, different frequency domains have been identified and associated with distinct phenomena, often described using terms such as breathing, flapping, and shedding.

The presence of a dominant frequency in TSBs, when scaled appropriately, was first recognised by \citet{Mabey1972}, who found that the separation bubble length $L_b$ and free-stream velocity $u_{\infty}$ provide a meaningful Strouhal number scaling, $St = f L_b / u_{\infty}$. Many TSBs exhibit a characteristic shedding frequency range between $St = 0.5$ and $St = 0.8$, attributed to vortex roll-up in the shear layer \citep{EatonJohnston1982, Kiya_Sasaki_1983, CherryHillier_1984_JFM}, often linked to the Kelvin–Helmholtz instability \citep{Tenaud2016}. \citet{Weiss2015UnsteadyBubble} observed a shedding mode at $St \approx 0.35$, associated with vortex roll-up in the shear layer above the recirculation region.

Beyond shedding, lower-frequency dynamics have also been observed. \citeauthor{EatonJohnston1982} noted already in \citeyear{EatonJohnston1982} that over 30\% of turbulence intensity occurred at frequencies below $St \approx 0.02$, though their measurement setup lacked the resolution to analyze this further. \citet{Kiya_Sasaki_1983} distinguished between high-frequency shedding and lower-frequency oscillations, describing the latter as an accumulation of vorticity that enlarges the bubble before being advected downstream.

In recent decades, two primary descriptions have become established for describing low-frequency motions: flapping and breathing. Flapping refers to oscillations of the shear layer, which have been reported in geometry-induced separation (such as blunt-plate configurations \citep{Kiya_Sasaki_1983}, backward-facing steps, etc.) and have also been discussed in forward-facing step configurations between $St \approx 0.08$ and $St \approx 0.18$ \citep{FangWang2024_JfM,Pearson2013,Largeau2006}.
Breathing, on the other hand, describes an overall expansion and contraction of the separation bubble, modulating both the separation and reattachment points. \citet{Pearson2013} characterised flapping as a streamwise and wall-normal expansion/contraction of the TSB, which \citet{WandGhaemi2022_JfM} argued aligns with the definition of breathing. While terminology varies, it is widely agreed that flapping occurs at intermediate frequencies, while breathing corresponds to the lowest frequencies, around $St \approx 0.01$ \citep{Weiss2015UnsteadyBubble, MohammedTaifur_Weiss_2016, Borgmann2024}.

The mechanism driving the breathing dynamics is still an open question. \citet{Wu2019} used a direct numerical simulation (DNS) to study a flat-plate TSB with two setups: one using suction-only separation and another using a suction-and-blowing profile to force reattachment. Dynamic mode decomposition (DMD) revealed streamwise-elongated structures potentially linked to Görtler instability, but their observed breathing frequencies ($St \approx 0.45$) were significantly higher than those reported in other studies. \citet{Cura2024} performed global stability (LSA) and resolvent analyses (RA) on a DNS dataset from \citet{Coleman2018NumericalNumber}. Their global stability analysis found no unstable eigenmodes, but the least damped mode occurred at zero frequency with small, nonzero spanwise wavenumbers. Resolvent analysis showed strong agreement with experimental databases, suggesting that forced dynamics are influenced by a weakly damped global mode. More recently, \citet{Cura2025} studied a family of turbulent separation bubbles of increasing size and showed that the growth rates scale with bubble dimensions, quantified by the reverse-flow magnitude and separation length. Unstable global modes were identified for the largest TSB extent. Another example in which a stationary three-dimensional mode has been identified in TSB research is provided by studies of stall cells. \citet{SarrasJfM2024} analyse stall cells using global linear stability analysis on a RANS mean flow and compare their findings with wind tunnel experiments. Their findings reveal stationary three-dimensional global modes becoming unstable only within certain ranges of angles of attack and particular spanwise wavenumbers.

The role of stationary three-dimensional eigenmodes as a driving feature of separation-bubble dynamics has already been demonstrated for laminar separation bubbles. \citet{Theofilis2000} analysed a laminar flat-plate flow and reported unstable, stationary, spanwise-periodic three-dimensional eigenmodes. \citet{Rodrguez2010} identified the same global mode and showed that it induces a spanwise modulation of the flow. Subsequently, \citet{Rodrguez2013} attributed this behaviour to a three-dimensional centrifugal instability and demonstrated that the associated growth rates depend on the reverse-flow intensity and the separation-bubble length.
 
\citet{Gallaire2007} similarly analysed a laminar separation bubble developing over a bump at $Re = 400$ and identified stationary three-dimensional unstable eigenmodes. Their study further examined whether these modes were associated with a centrifugal-type instability.
\citet{Marquet2009DirectNon} examined a laminar separation bubble behind a smooth ramp and found that destabilization occurred due to a stationary three-dimensional mode with a spanwise wavenumber. This mode was linked to the formation of high- and low-speed streaks, which they identified as a lift-up effect.

In the transitional regime, \citet{SavarinoJFM2025} investigated a flat-plate separation-bubble flow using both linear stability and resolvent analyses. They identified zero-frequency eigenmodes consistent with a centrifugal mechanism, while the corresponding resolvent spectrum exhibited a secondary hump at higher spanwise wavenumbers, that is related to a non-modal lift-up process. Additionally, they reported a configuration in which both centrifugal and lift-up mechanisms appeared to be active simultaneously.

\citet{Bugeat2022JfM} investigated the low-frequency breathing dynamics of a laminar oblique SBLI and conducted a RA on a steady 2D base flow. They found a modal mechanism for large spanwise wavelengths of the order of the separation bubble length with large gain values. At smaller spanwise wavelengths, in the order of the boundary layer thickness, they identified the growth of streaks, which they attributed to a non-modal mechanism.

Although \citet{Bugeat2022JfM}, \citet{Marquet2009DirectNon}, \citet{SavarinoJFM2025}, \citet{Gallaire2007}, \citet{SarrasJfM2024}, and \citet{Cura2024} investigate different flow configurations, their findings consistently indicate that the amplification behaviour of the low-frequency breathing mechanism in separation bubbles exhibits a clear spanwise dependency.
Given the strong influence of spanwise wavenumber on amplification and stability mechanisms, the spanwise dimension and boundary conditions likely play a crucial role. The impact of sidewalls and the distinction between periodic and solid-wall boundary conditions on low-frequency modes have been explored in SBLI research \citep{Rabey2019JfM,Garnier2009}, where they were attributed to corner effects. Similarly, in transitional backward-facing step configurations, \citet{Kaiktsis1991} and \citet{Barkley2002} reported that the type (solid vs. periodic) and spacing of sidewalls influences the spanwise modulation of the separation length. However, such influence remains largely unexamined in subsonic TSB studies. Many numerical studies \citep{Cura2024, Wu2019} utilise periodic sidewalls, as they minimise spurious boundary effects, reduce computational costs, and preserve statistical homogeneity. However, periodic boundary conditions impose constraints by allowing only integer spanwise wavenumbers. Experimental setups, in contrast, typically have solid sidewalls, that influence the flow dynamics.

The impact of spanwise boundary conditions on the low-frequency breathing dynamics of pressure-gradient-induced separation bubbles has not yet been explicitly investigated. It is often implicitly assumed that numerical studies with periodic sidewalls and experiments with solid sidewalls exhibit the same dynamics, making them directly comparable. The aim of this study is to examine the mode selection mechanisms governing low-frequency dynamics in experimental setups with solid boundaries. Additionally, we investigate the influence of periodic versus solid-wall boundaries on low-frequency dynamics and assess the role of the spanwise domain length in the mode selection process.

To address these questions, we conduct experiments on a turbulent separation bubble and perform particle image velocimetry (PIV) to obtain time-resolved planar velocity fields. We analyse the data using spectral proper orthogonal decomposition (SPOD) to extract coherent structures. Next, we perform LSA and RA on a 2D mean flow, generated via Reynolds-Averaged Navier-Stokes (RANS) simulations. By varying the spanwise wavenumber, we obtained solutions that are harmonic in the spanwise direction, representing three-dimensional traveling waves. We then construct a standing wave model by considering the sidewalls. Finally, by comparing the standing wave model with SPOD results from the experiment with fixed sidewalls, we validate our modeling approach. To further investigate the driving physical mechanisms of the breathing motion of the TSB, we analyze the response structures obtained from the RA.

\section{Acquisition and processing of experimental data}
\subsection{Experimental setup} \label{sec:experimental_setup}

\begin{figure}
\centering
\includegraphics[width=0.7\columnwidth]{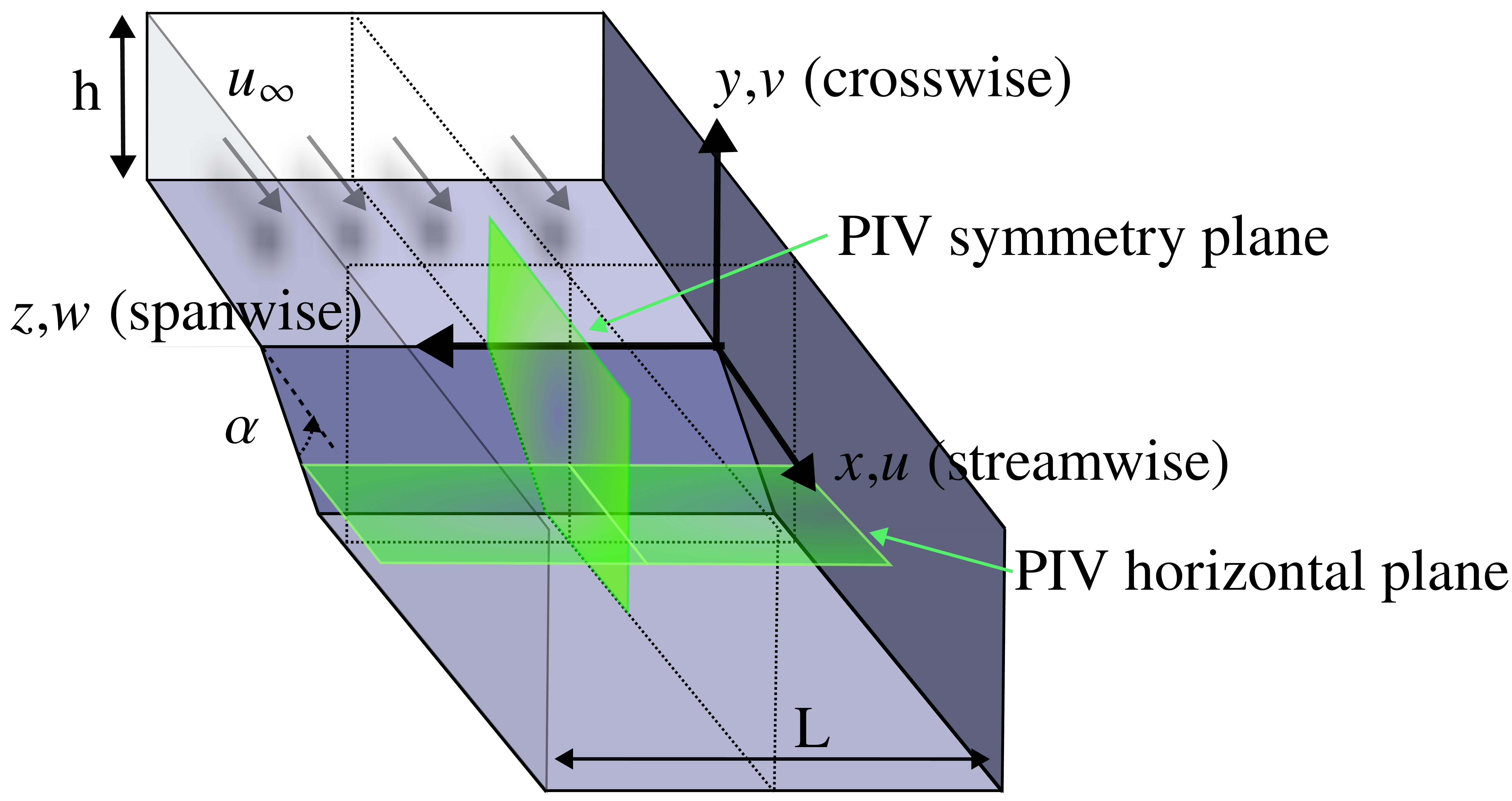}
\caption{Visualization of geometric setup, coordinate system and PIV planes.}
\label{fig:setup}
\end{figure}

The experiments were conducted in a low-speed closed-loop wind tunnel which maintains a constant Reynolds number throughout the measurements through temperature control.
The test section and the coordinate system are shown in Figure~\ref{fig:setup}. The free stream velocity of the test section is set to $u_\infty = 20\,\mathrm{m/s}$. The momentum thickness $\vartheta\approx 0.7\,\mathrm{mm}$ is determined using hot-wire anemometry at $x = -200\,\mathrm{mm}$, yielding a Reynolds number of $Re_\vartheta \approx 1\ 000$. Upstream of the ramp, the channel has a height of $h=400\,\mathrm{mm}$. The domain has a spanwise width of $L=600\,\mathrm{mm}$. Flow separation inside the test section was induced by an adverse pressure gradient due to the linear widening of the cross-section shown in Figure~\ref{fig:setup}. Reattachment occurs downstream of the backward-facing ramp, which was tilted by an angle of $\alpha=$ 20 degrees, as the separated shear-layer impinges onto the horizontal plate. The average streamwise length of the separation bubble on the centre line is approximately $L_b = 200\,\mathrm{mm}$. This characteristic length scale is used to normalise the frequency (using $St = f L_b / u_\infty$) and spanwise wavenumbers $\beta$ of the coherent structures throughout this article. Since the key focus of this study is to examine how sidewalls and spanwise domain length influence mode selection and the geometric scaling of coherent structures, the spanwise domain length $L$ is additionally referenced in many figures.

To provide quantitative information on the dynamics of the flow field, monoscopic particle image velocimetry measurements were conducted inside the symmetry ($xy$) plane and inside a horizontal ($xz$) plane. For the former, flow field information was obtained in the range $x = 0, \dots , 420\,\mathrm{mm}$ spanning the entire region of flow separation and reattachment inside the symmetry plane. The horizontal plane covered the full spanwise extent of the test section (from $z=0$ to $600\, \mathrm{mm}$) and included the streamwise range from $x = 193$ to $325\,\mathrm{mm}$. The vertical position of this plane $y\approx -70\,\mathrm{mm}$ ($45\,\mathrm{mm}$ above the bottom wall) was carefully chosen such as to capture the low-frequency dynamics observed in a preliminary study \citep{Steinfurth_TSFP_204}.

For PIV measurements, $d_\mathrm{p}=1\,\mathrm{\mu m}$ aerosole particles were illuminated by a high-speed diode-pumped Nd:YLF laser (\textit{Litron LD30-527}) inside light sheets with maximum thicknesses of three millimeters. Double images were recorded by one-megapixel cameras (\textit{Phantom VEO 710L}, $1 280\times 800\,\mathrm{px}$, $20\,\mathrm{\mu m}$ pixel pitch). The acquisition frequency was set to $f_\mathrm{s}=200\,\mathrm{Hz}$ ($St_\mathrm{s}\approx 2$), which is sufficiently large to capture the medium-frequency dynamics associated with vortex shedding while at the same time allowing for long recording durations required to resolve the low-frequency content. Specifically, 17~000 snapshots were taken over the course of $t_\mathrm{s}=85\,\mathrm{s}$ for the $xy$ plane, and 24~965 snapshots ($t_\mathrm{s}\approx 125\,\mathrm{s}$) were recorded for the horizontal plane.

The snapshots were pre-processed by masking the light reflections on the diffuser surface and minimum-intensity background subtraction. Two-component velocity vectors were computed using multi-grid cross-correlation with final interrogation area sizes of $(\Delta x, \Delta y)=(7.0\, \mathrm{mm}\times 5.2\, \mathrm{mm})$ and $\Delta x = \Delta z=3.2\, \mathrm{mm}$ for the symmetry  and horizontal planes, respectively. By applying a 50\% overlap of interrogation areas, the vector pitch was halved. As a post-processing step, a universal outlier detection \citep{Westerweel2005} was applied to replace implausible vectors.
To ensure the spatial resolution mentioned above, the velocity fields were obtained simultaneously with two cameras inside the symmetry plane and three cameras in the horizontal plane with overlapping fields of view. For each timestep, snapshots were projected onto a common structured grid spanning the fields of view by means of cubic interpolation.

\begin{figure}
    \centering 
    \includegraphics[width=1.0\textwidth]{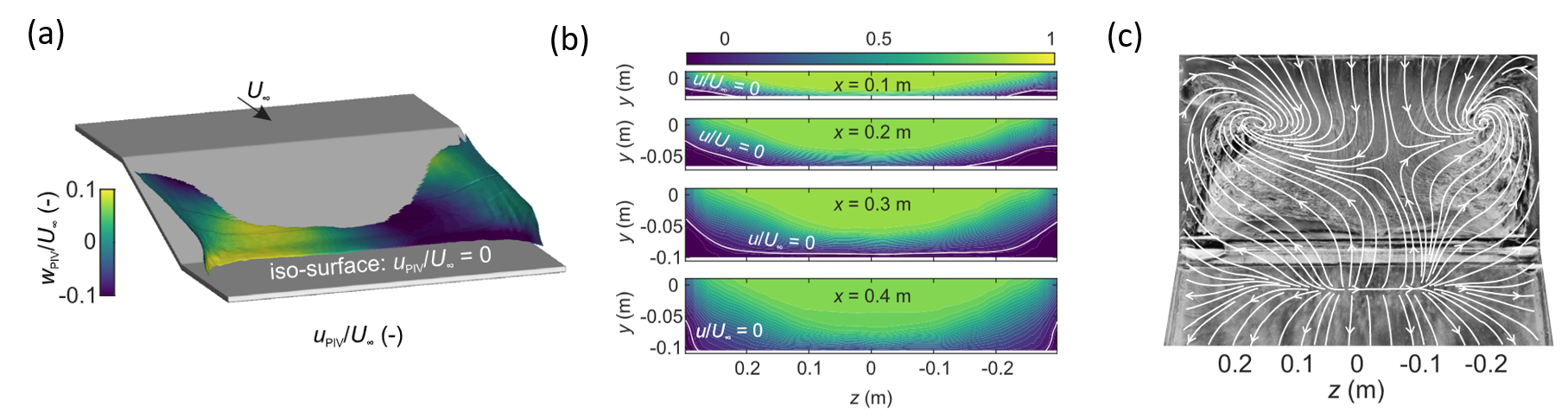}  
    \caption{Three-dimensional characteristics of the flow. (a) Iso-surface of the time-averaged streamwise velocity ($\bar{u} = 0$ m/s) from the 3D PINN model, representing the boundary of the turbulent separation bubble, coloured by the spanwise velocity. (b) Streamwise velocity component from the 3D PINN model at different cross-sections along the TSB. (c) Oil-film visualization overlaid with surface streamlines obtained from wall shear-stress measurements, where streamlines are reconstructed using bi-cubic interpolation of discrete measurement data. The Figures are published in \cite{SteinfurthPINN_PoF2024}.}
    \label{fig:3D_meanflow}
\end{figure}

It was recently shown by \cite{SteinfurthPINN_PoF2024} that the mean flow topology for this setup is three-dimensional, which is prominently manifested in a U-shaped mean separation line as seen in Figure \ref{fig:3D_meanflow}(a). They trained a physics-informed neural network (PINN) on pressure, wall-shear stress, and three-component velocity data obtained from PIV to reconstruct the 3D mean flow. Figure \ref{fig:3D_meanflow}(a) is based on this model and illustrates that flow separation occurs further upstream near the sidewalls, resulting in a longer separated region in these areas and a shorter separated region near the symmetry plane. The spanwise inhomogeneity of the mean flow is highlighted in Figure \ref{fig:3D_meanflow}(b), while the combination of wall shear stress data and oil-film images in Figure \ref{fig:3D_meanflow}(c) further emphasises the three-dimensional nature of the flow.

\subsection{Data processing using spectral proper orthogonal decomposition}

The PIV snapshots from both the symmetry and horizontal planes are analyzed using SPOD to extract structures of spatial and temporal coherence, referred to as coherent structures in the following.  

Proper orthogonal decomposition (POD) identifies dominant flow features by decomposing the velocity field into orthogonal spatial modes ranked by their energy content. While standard POD is purely spatial, SPOD extends this concept by performing a frequency-dependent decomposition. This allows the extraction of coherent structures associated with specific frequencies, providing insight into spectral flow dynamics.  

In this study, we apply the SPOD algorithm developed by \cite{Lumley1970, SchmidtColonius2020}. For the symmetry plane, both $u'(\boldsymbol{x}, t)$ and $v'(\boldsymbol{x}, t)$ velocity components are used as input. The time series is segmented into overlapping blocks of 200 snapshots with 50\% overlap. Each segment is windowed to reduce spectral leakage, and a Fourier transform is applied. 
The cross-spectral density (CSD) matrices are computed for each frequency based on the Fourier mode estimates of all blocks. Finally, an eigendecomposition of the CSD matrix yields the SPOD modes ranked by energy, analogous to standard POD but with spectral resolution as the CSD matrix is formed for each frequency separately.

A similar approach is applied to the horizontal plane PIV data. Here, we use $u'(\boldsymbol{x}, t)$ as input fields, with a window size of 200 snapshots and an overlap of 50 \%. The results of the SPOD analysis are presented in Chapter \ref{sec:Results}.

Later in the study, we compare SPOD modes with resolvent modes, following the approach of \citep{Towne2018}, who demonstrated that SPOD modes and RA modes are identical when the forcing is spatially uncorrelated. 

\section{Modelling Setup}

\subsection{RANS Setup} \label{sec:RANS}
Preliminary results from resolvent analysis reveal that the low-frequency modes considered in this work exhibit large streamwise wavelengths, which go far beyond the domain captured by the PIV measurements. To obtain a mean field that is sufficiently long to capture these dynamics via resolvent analysis it was necessary to conduct a RANS simulation and to tune it to match the PIV mean fields in the TSB region. 

\begin{figure}
    \centering
    \includegraphics[width=0.8\columnwidth]{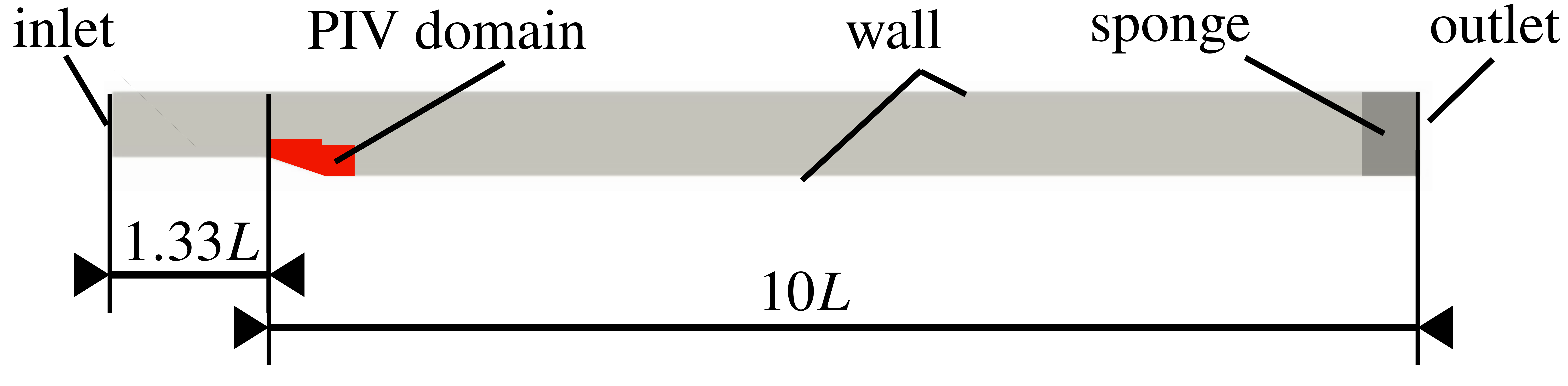}
    \caption{RANS domain, PIV symmetry plane domain, boundary names and sponge region}
    \label{fig:RANS_domain}
\end{figure}

Another advantage of using a RANS simulation is that it allows for the direct extraction of an eddy viscosity, which is necessary for the resolvent analysis \citep{Pickering2021,Kuhn2021,vonSaldern_JFM_2024} and will be discussed in detail in Section \ref{sec:resolvent}. The eddy viscosity field is smooth and readily obtainable from the RANS simulation. Although eddy viscosity could also be derived from Reynolds stresses \citep{Oberleithner2013c,Tammisola2016CoherentModes,Rukes2015} using the Boussinesq model and a least squares fit, this approach often results in noisy eddy viscosity fields \citep{Beresh2018PracticalData} and the problem of insufficient spatial coverage would remain.

The RANS simulation is performed with the open-source finite volume solver OpenFOAM (version 2212). The domain is discretized using a quasi-2D mesh that contains only one cell in the spanwise direction $z$. The mesh is a hybrid mesh that contains structured wall normal prism cells in the entire domain except for a region near the ramp where unstructured hexahedral cells are used. The mesh has 632~658 cells. The mesh has been chosen based on a mesh refinement study that included meshes with 21~085, 83~838, 331~555, 632~658 and 1~208~926 cells. 
The incompressible formulation of the RANS equations is implemented using the semi-implicit method for pressure linked equations algorithm (SIMPLE) \citep{PatankarSpalding_SIMPLE}. To model the eddy viscosity the common two-equation turbulence model k-$\omega$-SST \citep{kOmegaSST_baseModel, kOmegaSST_updated} is utilized. 

Figure \ref{fig:RANS_domain} shows the RANS domain, its boundaries and the spatial coverage of the available PIV data in the symmetry plane. The boundary conditions are defined as follows. At the inlet there is a uniform velocity inlet with $\boldsymbol{u}=[u_{\infty},0,0]^T$, $u_{\infty}=20\ m/s$  and $\p p /\p x = 0$ at $x=-0.8m$ ($-1.33L$), a pressure outlet is applied at $x=6m$ ($10L$) where the modified pressure is set to zero and the velocities have a zero gradient in wall normal direction. On the top and bottom surfaces, a no-slip wall condition is applied without using wall functions. The height $h$ of the numerical domain equals the height of the experimental setup. The cell sizes are set such that $y^+ < 1$ holds on all no-slip walls. 

\begin{figure}
    \centering
    \includegraphics[width=0.9\columnwidth]{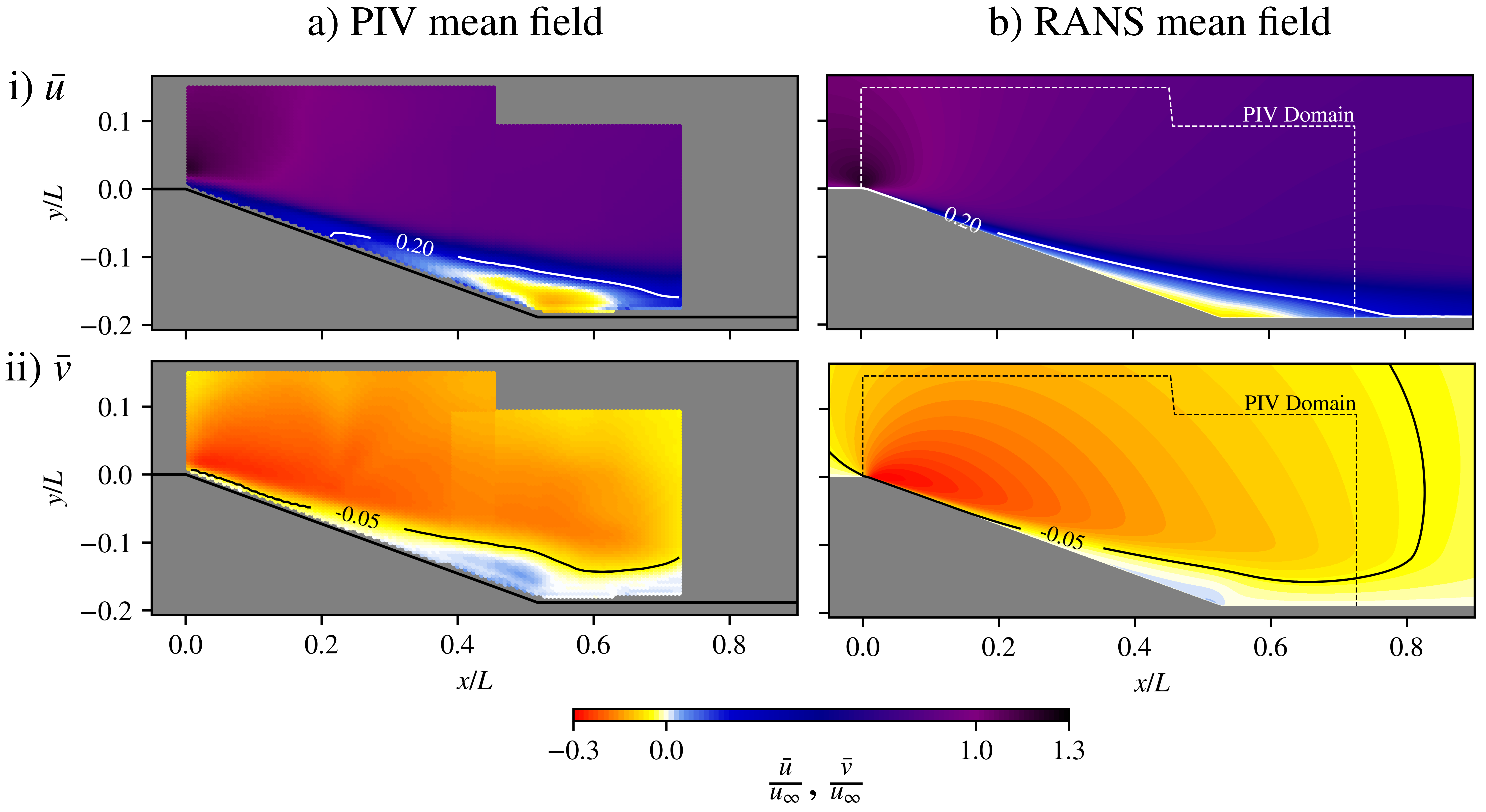}
    \caption{Comparison of mean velocity components from PIV measurements and RANS simulation. (a) PIV mean fields: (i) streamwise velocity component $\bar{u}$ with the $\bar{u}/u_\infty = 0.2$ iso-contour indicated by a white line; (ii) vertical velocity component $\bar{v}$ with the $\bar{v}/u_\infty = -0.05$ iso-contour shown as a black line. The geometry boundary is also visualized by a black line. (b) Corresponding RANS  mean fields, with the PIV measurement domain outlined by a dashed line. Coordinates are scaled with the spanwise domain length $L$.}
    \label{fig:RANS_mean_field}
\end{figure}

The initial condition for the turbulence kinetic energy $k$ and the turbulence specific dissipation rate $\omega$ can be set based on characteristic length and velocity scales and a turbulence intensity value \citep{openFoamUserGuide_kOmegaSST}. The latter describes the ratio between the root mean square of the turbulent velocity fluctuations and the mean velocity in percent. A value of 3.75\% was carefully chosen aiming to minimize the difference in separation bubble size between the RANS solution and the PIV mean field. This approach takes advantage of the strong influence of freestream turbulence on the separation point of the boundary layer from the surface. In particular, higher values of free stream turbulence cause a larger momentum transfer normal to the streamwise direction of the flow. This results in a longer capability of the flow to stay attached. So the separation point is moved further downstream, resulting in a smaller separation bubble. Lower freestream turbulence values move the separation point upstream, resulting in a larger separation bubble. 
This approach yields a similar inflow momentum thickness as in the PIV measurements. While at $x = -200\,\mathrm{mm}$, the RANS simulation indicates a larger value ($\vartheta \approx 1.1\,\mathrm{mm}$) compared to the experiment ($\vartheta \approx 0.7\,\mathrm{mm}$), the deviation is marginal at $x=-50\,\mathrm{mm}$ where $\vartheta \approx 0.9\,\mathrm{mm}$ (PIV) versus $\vartheta \approx 1.0\,\mathrm{mm}$ (RANS).
Figure \ref{fig:RANS_mean_field} shows the $\bar{u}$- and $\bar{v}$-components of the resulting RANS mean field alongside the PIV mean field.
Iso-contours of $\bar{u}/u_\infty = 0.2$ and $\bar{v}/u_\infty = -0.05$ are included to allow for a more quantitative comparison. While the flow topologies agree reasonably well, slight deviations can be observed in the recirculation region: the PIV data show a slightly larger separation bubble in the crosswise direction with  stronger reverse flow ($-u_{\text{rev}}/u_\infty = 13\%$ for PIV compared to $7\%$ for RANS, where $u_{\text{rev}}$ denotes the peak reverse flow of the mean field). The RANS mean field displays a thin separation bubble in the upstream part which is not visible in the PIV mean field.
This latter observation may be attributed to the presence of spurious measurement data near the wall in the current dataset, caused by low signal-to-noise ratios due to strong light reflections. This experimental shortcoming does not negatively affect the SPOD analysis, since the breathing mode unfolds further away from the wall, as will be shown later.
With the chosen RANS model using the $k$-$\omega$-SST turbulence model and adaption of the turbulence intensity value at the inlet of the domain, it was not possible to better approximate the crosswise extent of the bubble without significantly altering the separation and reattachment points. 
Similar challenges in matching RANS-predicted separation bubble behaviour with experimental observations have been reported for smooth-body separation, where accurate prediction remains difficult and continues to be an active topic of research \citep{williams_Experimental_2020, gray_Experimental_2023,Klopsch2025}. \citet{SarrasJfM2024} observed comparable discrepancies and showed that improved agreement could be achieved by applying adjoint-based data assimilation with flexibility in the turbulence model.

In the present setup, as shown in Figure~\ref{fig:RANS_mean_field}, the RANS simulation approximates the PIV data well despite the deviations discussed above. Considering that the PIV field reflects a fully three-dimensional flow, whereas the RANS solution is based on a quasi-two-dimensional setup, the level of agreement is very satisfactory. Further validation will be provided in Section~\ref{sec:Results} through a comparison between data-driven SPOD modes and resolvent modes computed from the RANS mean field.

\subsection{Global Linear Stability Analysis} \label{sec:LSA}
To investigate modal amplification mechanisms in the flow, a global linear stability analysis (LSA) of the turbulent mean flow is performed. This approach examines the time-asymptotic dynamics of harmonic perturbations that evolve around the temporal mean state by performing an eigenvalue decomposition of the linearized Navier--Stokes operator, which yields growth rates and corresponding mode shapes. Linear stability analysis has become a common framework to study coherent flow dynamics \citep{Sipp2010,Barkley2006} and has recently been applied to separation bubble flows \citep{Cura2024,Cura2025,SarrasJfM2024}.

Due to low Mach numbers ($Ma < 0.1$) the incompressible formulation of the Navier-Stokes equations is used. The state vector $\boldsymbol{q}(\boldsymbol{x}, t) = [\boldsymbol{u}, p]^T$ consist of the velocity vector $\boldsymbol{u}=[u,v,w]^T$ and the pressure $p$ and are defined on Cartesian coordinates $\boldsymbol{x}=[x,y,z]^T$.

To derive the linearized operator, the Reynolds decomposition is substituted into the Navier--Stokes equations. The Reynolds decomposition is applied to the state vector $\boldsymbol{q} = \boldsymbol{\Bar{q}}+\boldsymbol{q'}$, where $\boldsymbol{\Bar{q}}$ describes the time-average and $\boldsymbol{q'}$ the coherent fluctuations. Subsequently, the time average is subtracted, resulting in the transport and continuity equation of the fluctuating velocity component, given by

\begin{equation}
\begin{split}
      \frac{\p \boldsymbol{u}'}{\p t}= -(\overline{\boldsymbol{u} }\bcdot \bnabla) \boldsymbol{u}' - (\boldsymbol{u}' \bcdot \bnabla) \overline{\boldsymbol{u} } - \frac{1}{\overline{\rho}}\bnabla p' + \bnabla \bcdot \Bigl[ (\nu + \nu_t)  [\bnabla+\bnabla^T] \boldsymbol{u}' \Bigl]  \\
      \bnabla \cdot \boldsymbol{u}' = 0
       \label{eq:coherent_full} \ .
\end{split}
\end{equation}

The variable $\rho$ denotes the mass density and $\nu$ the kinematic viscosity. Expanding all terms in Eq.~\ref{eq:coherent_full} explicitly reveals the turbulent forces $\bnabla \bcdot (\boldsymbol{u}' \boldsymbol{u}')$, which represents an unknown term in the analysis and is modelled using a Boussinesq-type turbulence model. The turbulence closure model relates the unknown forces via an eddy viscosity $\nu_t$ to the fluctuating strain rate tensor. 
This ad hoc modeling choice, analogous to turbulence modeling in the Reynolds-averaged Navier-Stokes equations, is known to improve the results for analyses based on linearized Navier--Stokes equations, as it introduces a turbulent dissipation mechanism into the analysis~\citep{Symon2023UseFlow,vonSaldern_JFM_2024}.
As eddy viscosity, the spatially non-uniform eddy viscosity field from the RANS simulation $\nu_t =\gamma \ \nu_{t,\text{RANS}}(x,y)$ is applied, where $\gamma$ denotes a scaling factor. To apply the eddy viscosity field from the RANS solution in the linearized analysis is a common approach for example used by \cite{Rukes2015, Yim2019, Pickering2021, Busquet_Marquet_JfM_2021,Klopsch2025}. The choice of the scaling factor is detailed in Section~\ref{sec:LSA_results}, and has been chosen to match the low-pass filter behaviour found experimentally.

The linear system is then decomposed into a Fourier basis. We assume the perturbations to be harmonic in time and in the spanwise direction $z$, which allows the modal ansatz
\begin{align}
\boldsymbol{q}'(x,y,z,t) = \boldsymbol{\Hat{q}}(x,y) \ e^{i(\beta z - \omega t)} + c.c , \label{eq:ansatz_LSA}
\end{align}
where $i$ is the imaginary unit, $\beta$ is the spanwise wavenumber, and $\omega = \omega_r + i\omega_i$ is the complex frequency, with $\omega_r$ representing the oscillation frequency, $\omega_i$ the temporal growth rate and $c.c$ the complex conjugate. Substituting (\ref{eq:ansatz_LSA}) into the linearized equation (Eq.~\ref{eq:coherent_full}) leads to the generalized eigenvalue problem
\begin{align}
(\boldsymbol{A}-i\omega \boldsymbol{B}) \boldsymbol{\Hat{q}} = 0, \label{eq:LSA_evp}
\end{align}
where $\boldsymbol{A}$ denotes the discretized linearized Navier--Stokes operator, including the continuity coupling, and $\boldsymbol{B}$ is the mass matrix arising from the spatial discretization. Solving this eigenvalue problem yields a spectrum of complex eigenvalues $\omega$ and eigenvectors $\boldsymbol{\Hat{q}}$. Eigenvalues with $\omega_i > 0$ correspond to exponentially growing (so called) unstable modes, while $\omega_i < 0$ indicates stable, decaying modes.

\subsection{Resolvent Analysis} \label{sec:resolvent}
Complementary to the global stability analysis, we perform a resolvent analysis. Unlike linear stability analysis, resolvent analysis includes non-modal amplification mechanisms and allows modeling coherent structures with broadband dynamics. It identifies the most amplified linear response to an external forcing in a time-invariant flow. Initially, it was used in stability and transition studies \citep{FarrellIoannou1993, trefethen199, SchmidHenningson2001}. Later \cite{McKeonSharma2010} showed that resolvent analysis can be applied to the mean of a turbulent flow. In recent years, it was applied in numerous studies to turbulent mean flows to uncover dominant physical mechanisms that form large-scale coherent structures \citep{Illingworth_2018,Pickering_LiftUp2020, Schmidt2018JfM, MonsMarquet2024_DA, Mueller_2024JfM}. 

For the resolvent analysis, we consider the linearized equations formulated in Eq.~\ref{eq:LSA_evp} but with an external forcing term $\boldsymbol{\Hat{f}}$,
\begin{align}
(\boldsymbol{A}-i\omega \boldsymbol{B})\boldsymbol{\Hat{q}} = \boldsymbol{D}\boldsymbol{\Hat{f}} , \label{eq:RA_f}
\end{align}
where $\boldsymbol{D}$ is the input (forcing) matrix that restricts the forcing to the momentum equations.
Equation~\ref{eq:RA_f} can be rearranged in an input-output form
\begin{equation}
    \hat{\boldsymbol{q}} = \boldsymbol{R}(\omega, \beta) \hat{\boldsymbol{f}} 
\end{equation}
where $\boldsymbol{R}=(\boldsymbol{A}-i\omega \boldsymbol{B})^{-1} \boldsymbol{D}$ denotes the resolvent operator which establishes a linear relation between the forcing $\hat{\boldsymbol{f}}$ and response $\hat{\boldsymbol{q}}$ at frequency $\omega$ and spanwise wavenumber $\beta$.

A singular value decomposition of the operator, $\mathcal{R} = \boldsymbol{U}\boldsymbol{\Sigma}\boldsymbol{V}^*$, at the selected frequency $\omega$ and spanwise wave number $\beta$ yields pairs of optimal forcing and response modes contained in $\boldsymbol{V}$ and $\boldsymbol{U}$, respectively. The diagonal matrix $\boldsymbol{\Sigma}$ contains the corresponding resolvent gain values that represent the amplification between forcing and response. The resolvent response mode associated with the largest gain value represents the coherent structure that is most amplified by linear mechanisms in the flow. As discussed above, these optimal response modes closely approximate the dominant structures extracted by SPOD when the true turbulent forcing is uncorrelated~\citep{Towne2018}.

\subsection{Numerical Setup for Stability Analysis and Resolvent Analysis}

We solve for the linear stability eigenmodes and the resolvent modes using the in-house software FELiCS, which uses a finite element discretisation scheme~\citep{Kaiser_felics}. The computational mesh has the same dimensions as the mesh used for the RANS simulation, but it is a fully unstructured mesh consisting only of triangles with a total of 33 222 cells. It was generated using the open source software Gmsh \citep{gmsh}. 
The boundary conditions for $\hat{u}, \hat{v}, \hat{w}$, and $\hat{p}$ are set to zero on all boundaries, which corresponds to a Dirichlet condition. The only exception is the pressure $\hat{p}$ at the wall, where a zero-gradient condition, known as a homogeneous Neumann condition, is applied.

To avoid spurious pressure modes near the inlet and the outlet a local sponge region is added. The region causes the mode to decay before reaching the boundary. This is implemented through a dissipative term in the momentum balance that only acts in the sponge region. It primarily serves to absorb and minimize reflections from computational boundaries, ensuring that artificial reflections do not interfere with the simulation results \citep{Bodony2006}. By varying the sponge region we ensured that it does not change the upstream modes and only dampens spurious modes near the outlet. 

For computing the LSA, FELiCS employs a SLEPc-based solver to obtain the eigenvalues and eigenvectors of the discretized problem. The solver returns a specified number of eigenvalues located nearest to a user-prescribed initial guess, along with their corresponding eigenmodes.

The forcing and response domain in the resolvent analysis, which can be freely chosen by adding a restriction operator similar to \cite{Towne2018}, was set to cover the entire simulation domain, except for the near-outlet region where the sponge layer is applied. This exclusion prevents the sponge from influencing the gain by altering the response modes. 

\subsection{Standing Wave Model} \label{sec:standing_wave_model}

Previous studies have shown that the mean flow under consideration has inherently three-dimensional characteristics \citep{SteinfurthPINN_PoF2024}. Ideally, a fully three-dimensional resolvent analysis conducted on a three-dimensional mean flow would provide the most accurate representation. However, such an approach introduces significant computational and numerical challenges and complicates direct comparisons with recent 2D resolvent studies \citep{Cura2024}. 
To address these limitations, a quasi-3D resolvent ansatz is introduced in the following that is based on a 2D mean flow and models fluctuations as 3D harmonic waves in the spanwise direction, $z$. Beyond its computational advantages, this formulation allows for an investigation into whether the linear amplification behaviour of the modes is primarily dictated by the 2D mean flow and how the spanwise boundary conditions influence the resulting dynamics. Such an analysis would not be easily achievable within a fully 3D resolvent framework.
 
In the following section, we systematically examine the constraints imposed by different spanwise boundary conditions on the possible solutions and develop a so-called {\itshape standing wave model} that appropriately incorporates these physical constraints.

The ansatz used for the linear stability and resolvent analysis in Eq.~\ref{eq:ansatz_LSA} can be more explicitly written as

\begin{align}
{\boldsymbol{q}'(\boldsymbol{x},t)}=
 \begin{pmatrix}
u'(x,y,z) \\
v'(x,y,z) \\
w'(x,y,z) \\
p'(x,y,z) \\
\end{pmatrix}
=
\begin{pmatrix}
\hat{u}(x,y) \\
\hat{v}(x,y) \\
\hat{w}(x,y) \\
\hat{p}(x,y) \\
\end{pmatrix}
 e^{i(\beta z-\omega t)} + c.c \label{eq:travelling_wave}
\end{align}

\begin{figure}
\centering
\includegraphics[width=1.0\linewidth]{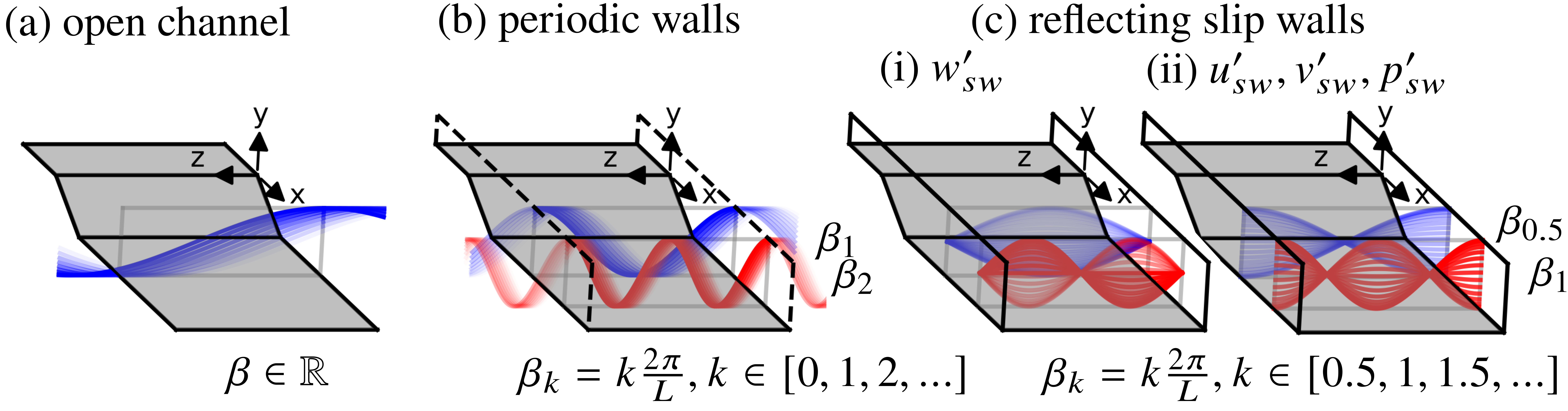}
\caption{Ramp geometry with spanwise wave visualizations. The wavenumber of each example wave is highlighted, following $\beta_k = k\frac{2\pi}{L}$. (a) Open infinite channel with a traveling wave of $\lambda_z > L$. (b) Periodic walls, showing the two largest possible waves: $\beta_1 = \frac{2\pi}{L}$ (blue) and $\beta_2 = \frac{4\pi}{L}$ (red). (c) Reflecting slip walls: (i) spanwise component $w'_{sw}$ and (ii) other components ($u'_{sw},\ v'_{sw},\ p'_{sw}$), illustrating the smallest possible wavenumbers, $\beta_{0.5} = \frac{\pi}{L}$ (blue) and $\beta_1 = \frac{2\pi}{L}$ (red).}
\label{fig:waves}
\end{figure}

This ansatz represents a wave with a continuous spanwise wavenumber $\beta \in \mathbb{R}$ and the corresponding spanwise wavelength $\lambda_z = 2\pi/\beta$. The longest possible wavelength $\lambda_{z, \text{max}}$ with this ansatz is unbound: $\lambda_{z, \text{max}}=\infty$. Without any restrictions, this model resembles a travelling wave in an open (infinite) channel, which is visualized in Figure \ref{fig:waves}(a). The wave can travel in both positive and negative $z$ direction. The sign of $\beta$ determines which direction the wave is travelling (when only positive frequencies are assumed).

Many numerical studies employ periodic sidewalls (\cite{Wu2019, Cura2024, FangWang2024_JfM}) instead of open (infinite) boundaries. Periodic boundary conditions are often preferred because they minimize spurious boundary effects. Additionally, they reduce computational costs by allowing for smaller domains while approximately preserving the statistical homogeneity of turbulence. However, periodic sidewalls impose an artificial constraint by enforcing a strictly periodic solution. As a result the computed pressure and velocity fields are also periodic in spanwise direction. This means that only fluctuations with integer spanwise wavenumbers are permitted, given by $\beta_n = n \frac{2\pi}{L}$ with $n \in \mathbb{Z}$. Accordingly, the largest finite wavelength is constrained by the channel width $L$, such that $\lambda_{z, \text{max}} = L$ (for $n=0$, the corresponding wavelength is infinite $\lambda_{z, \text{max}} = \infty$).
For this setup, waves can propagate in both positive and negative spanwise directions, depending on the sign of the wavenumber. An example of this configuration with the two largest allowable waves excluding $n=0$ ($\beta_1=\frac{2\pi}{L}$ and $\beta_2=\frac{4\pi}{L}$) is illustrated in Fig. \ref{fig:waves}(b).

In the experimental setup described in Chapter~\ref{sec:experimental_setup} the side walls are solid. As with the periodic setup, the wavelength is constrained by the channel width, $L$, and the wavenumbers are discrete. The presence of a solid wall, however, introduces an additional constraint.
In solid mechanics and acoustics, wave reflection at boundaries is a well-known phenomenon. A simple demonstration is the classic experiment where a wave pulse traveling along a string fixed at one end reflects upon reaching the boundary. Similarly, water waves hitting solid surfaces exhibit reflection, and in acoustics, echoes in large spaces provide an intuitive example of wave reflection. In compressible numerical simulations, boundary conditions incorporating reflection and refraction properties are standard. However, while our setup is incompressible, we still assume reflection of hydrodynamic waves at the walls.

To include this in our resolvent model, we adopt a simplified assumption: flow perturbations are reflected upon encountering a boundary. We draw an analogy to turbulent bands propagating in high-aspect-ratio channel flows. \cite{Kohyama2022} demonstrate using DNS, that turbulent bands, which characterize the spatial intermittency of laminar-turbulent interfaces, are convected by the bulk flow and reflect upon reaching spanwise sidewalls. By analogy, we assume that linear perturbations in the resolvent framework behave similarly -- reflecting at solid sidewalls. Thus, we model the walls as reflective boundaries for resolvent modes. This is the basis for our standing-wave model.

To derive the standing-wave model, we consider a wave with an arbitrary wavenumber $\beta$ that travels in positive spanwise direction and is reflected at the wall at $z = L$, undergoes a change in orientation that changes the sign of the wavenumber, and propagates to $z = 0$, where it is reflected again. In the presence of reflecting boundaries at both ends, a wave is always accompanied by a reflected wave with the same wavelength but inverted orientation. The superposition of two waves of opposite directions with the same wavelength results in the formation of a standing wave.
The ansatz for the standing-wave model is then given as:
\begin{align} 
\boldsymbol{q}'_{\text{sw}}(\boldsymbol{x},t)=\boldsymbol{q}'_{+}(\boldsymbol{x},t)+\boldsymbol{q}'_{-}(\boldsymbol{x},t) \label{eq:standing_wave1}
\end{align} 
where the subscript $(\cdot)_{\text{sw}}$ denotes a standing wave and $(\cdot)_{+}$ a traveling wave that is defined by Eq. \ref{eq:travelling_wave} that travels in positive $z$ direction and $(\cdot)_{-}$ in negative $z$ direction. In Eq.~\ref{eq:standing_wave1}, $\boldsymbol{q'_{+}}$ and $\boldsymbol{q'_{-}}$ correspond to left- and right-traveling resolvent modes for the same absolute wavenumber $\beta$ but with opposite signs. Due to symmetry considerations, a resolvent mode at frequency $\omega$ and wavenumber $\beta$ is equivalent to the resolvent mode at the same frequency and same but negative wavenumber and $w$-component. Utilizing this symmetry, we can formulate the standing-wave model 
\begin{align}
\boldsymbol{q}'_{\text{sw}}(\boldsymbol{x},t)=
\underbrace{
\begin{pmatrix}
\hat{u}(x,y) \\
\hat{v}(x,y) \\
\hat{w}(x,y) \\
\hat{p}(x,y) \\
\end{pmatrix}
 e^{i(\beta z-\omega t)}
 }_{\mathcal{\boldsymbol{R}}(\omega, \beta) \ \rightarrow \ \boldsymbol{q'_{\text{+}}} }
 +
 \underbrace{
\begin{pmatrix}
\hat{u}(x,y) \\
\hat{v}(x,y) \\
-\hat{w}(x,y) \\
\hat{p}(x,y) \\
\end{pmatrix}
 e^{i(-\beta z-\omega t)} 
  }_{\mathcal{\boldsymbol{R}}(\omega, -\beta) \ \rightarrow \ \boldsymbol{q'_{\text{-}}} }
  + c.c
\label{eq:standing_wave}
\end{align}
that is constructed of two resolvent modes with opposite spanwise phase velocities. The same approach could be applied to the eigenmodes.
To incorporate spanwise boundary conditions, we enforce slip walls. The slip-wall boundary condition constrains the wall-normal velocity perturbation, $w'_{sw}$, to zero at $z=0$ and $z=L$. This holds for all spanwise wavenumbers $ \beta$  at $z=0$, but only for $\beta = n\pi/L$ , $n \in \mathbb{Z}$, at $z=L$. The allowed wavenumbers in the standing wave model are thus  
\begin{equation}
\beta_k = k \frac{2\pi}{L}, \quad k \in  
\begin{cases} 
    0, 1, 2, \dots & \text{for periodic sidewalls,} \\
    0, 0.5, 1, 1.5, \dots & \text{for slip walls.}
\end{cases}
\end{equation}
Therefore, the longest wavelength for a reflective slip-wall condition is \( \lambda_{z, \text{max}} = 2L \).  
This restriction also affects the other flow components. Applying it to the standing wave form of the streamwise velocity, we find that \( u'_{sw} \sim \cos(\beta_k z) \), where \( \beta_k = k\pi/L \). Consequently, the amplitude of \( u'_{sw} \) is always maximal at \( z=0 \) and either maximal or minimal at \( z=L \), but never zero. Thus, the slip-wall condition forces nodes at the walls for \( w'_{sw} \) and antinodes for \( u'_{sw} \), \( v'_{sw} \), and \( p'_{sw} \). Hence, while this analysis remains two-dimensional in its essence, the spanwise direction is treated as a periodic continuation. This quasi-3D ansatz ensures that each individual wave develops on a 2D mean flow and is not directly influenced by 3D sidewall effects.

Note that with this quasi-3D ansatz, the more realistic no-slip boundary condition at the sidewalls cannot be satisfied. This is due to the relative phase positioning between the \( u'_{sw} \), \( v'_{sw} \), and \( w'_{sw} \) components given by the resolvent modes. Figure \ref{fig:waves}(c) illustrates this behaviour for the two largest standing waves, $ \beta_{0.5} = \frac{\pi}{L} $ and $ \beta_1 = \frac{2\pi}{L} $. Panel (i) and  (ii) visualize the axial and spanwise component of the standing-wave fluctuations clearly demonstrating that the one attains its maximum precisely when the other vanishes, and vice versa.
To enforce a no-slip boundary condition at the sidewalls while avoiding the trivial solution, a fully three-dimensional resolvent analysis of the 3D mean flow would be required. However, this approach introduces additional complexity, as it would necessitate considering inherently 3D mean flow phenomena, making the interpretation of results more challenging.

By adopting our quasi-3D model with reflecting slip walls that generate standing waves, we do not only significantly reduce the complexity compared to a full 3D resolvent model. In fact, we can investigate whether the amplification behaviour of the flow dynamics is primarily a 2D mean flow effect that is secondarily dominated by the spanwise boundary. This would not be as easy with a 3D resolvent analysis.

In summary, boundary conditions impose fundamental restrictions on the allowable wavelengths. Periodic sidewalls, often employed in numerical studies, permit only integer wavenumbers, with the maximum wavelength constrained by the channel width. By drawing an analogy to acoustics and water waves, we assume the sidewalls exhibit a reflective property. In this framework, resolvent-based waves undergo reflection at the sidewalls, leading to the formation of standing waves. The validity of this model and its agreement with experimental data will be demonstrated in the following section.

\pagebreak

\section{Results} \label{sec:Results}
\subsection{Empirical description of the TSB low-frequency dynamics}

\begin{figure}
    \centering 
    \includegraphics[width=0.7\textwidth]{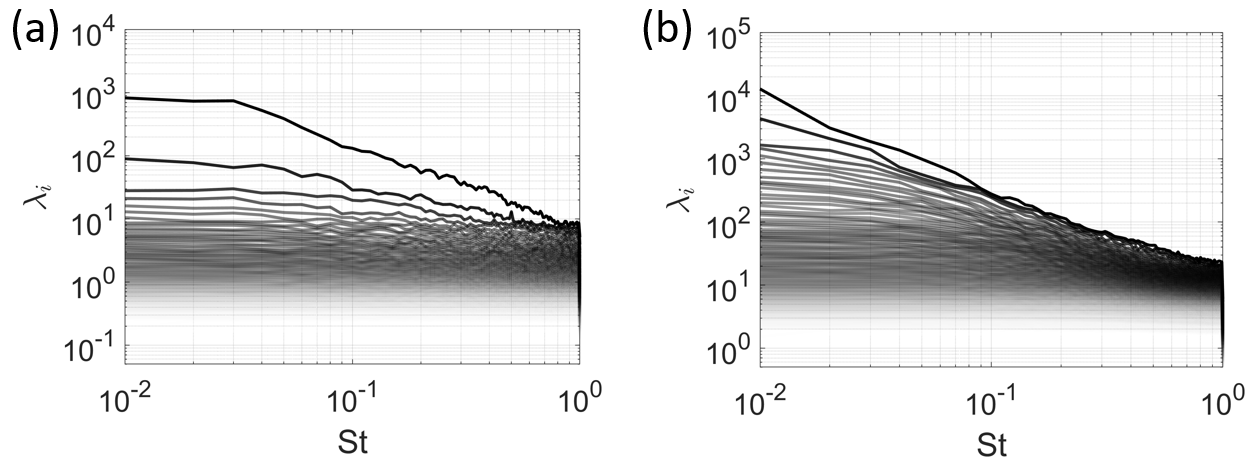}
    \caption{SPOD spectrum of (a) symmetry plane PIV dataset and (b) horizontal plane PIV dataset}
    \label{fig:SPOD_spectrum_combined_unnormed}
\end{figure}

\begin{figure}
    \centering 
    \includegraphics[width=0.7\textwidth]{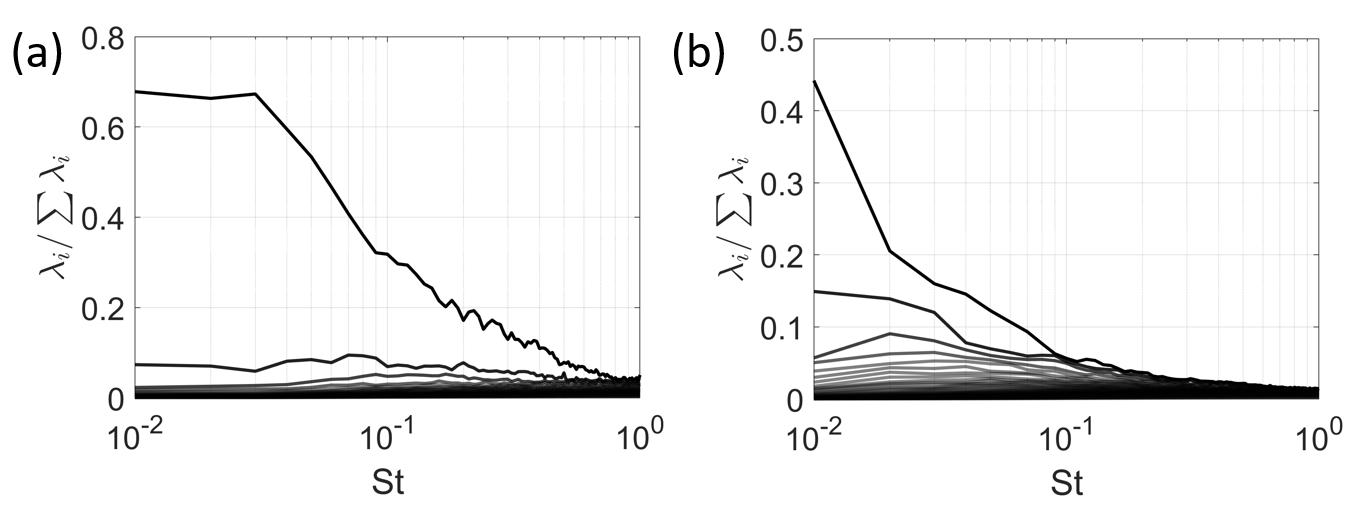}
    \caption{Energy share of the SPOD modes of (a) symmetry plane PIV dataset and (b) horizontal plane PIV dataset}
    \label{fig:SPOD_spectrum_combined}
\end{figure}

The SPOD spectrum of the streamwise velocity fluctuations measured by PIV on the symmetry plane is presented in Figure \ref{fig:SPOD_spectrum_combined_unnormed}(a) which shows the mode energies of all SPOD modes $\lambda_i$. Figure \ref{fig:SPOD_spectrum_combined}(a) shows the energy share of each mode at every frequency which is calculated by dividing the mode energy by the sum of all modes at the given frequency. It reveals a leading mode that has a significantly higher energy share compared to the other modes. This is also known as high-gain separation and describes low-rank dynamics, meaning that the dynamics of the flow are dominated by one or very few coherent structures in the corresponding frequency range \citep{SchmidtColonius2020}. 
This mode concentrates the majority of its energy within the low-frequency range, specifically below $St=0.05$. The spectral resolution of the SPOD is determined by the sampling frequency and the number of snapshots per block. In the symmetry plane, a window size of 200 snapshots yields a minimum resolvable frequency of $St = 0.01$. To resolve lower frequencies, we increased the window length to 2000 samples, extending the resolution down to $St = 0.001$ (not shown). While this confirmed the presence of the same low-pass filter behavior, the resulting spectra exhibited reduced statistical convergence due to the lower number of available blocks for averaging. A similar low-rank behaviour for low frequencies has been observed by \cite{WeissAIAAJ2022} using SPOD of wall-shear stress signals. Instead of a single dominant characteristic frequency, the spectrum exhibits a prominent band of low frequencies. 

If the spectrum were pre-multiplied, it would be forced to zero at zero frequency, and a distinct hump would appear between $St = 0.01$ and $St = 0.05$. This frequency range is typically associated with breathing when considering pre-multiplied spectra \citep{PoggieAIAA2015, MohammedTaifourJfM2021}. In the non–pre-multiplied representation, however, the response exhibits little frequency selectivity for $St \lesssim 0.05$. This is characteristic of a low-pass filter, which is a key feature of low-frequency breathing dynamics \citep{Cura2024,MohammedTaifourJfM2021}. This concept traces back to \cite{Plotkin1975}, who first proposed it in the context of shock oscillations in turbulent SBLI flows. \cite{PoggieAIAA2015} later demonstrated its relevance to a larger number of SBLI flows and \citet{MohammedTaifur_Weiss_2016, MohammedTaifourJfM2021} extended the concept to a low-speed TSB case which was also observed by \cite{Cura2024}.

\begin{figure}
    \centering 
    \includegraphics[width=0.99\textwidth]{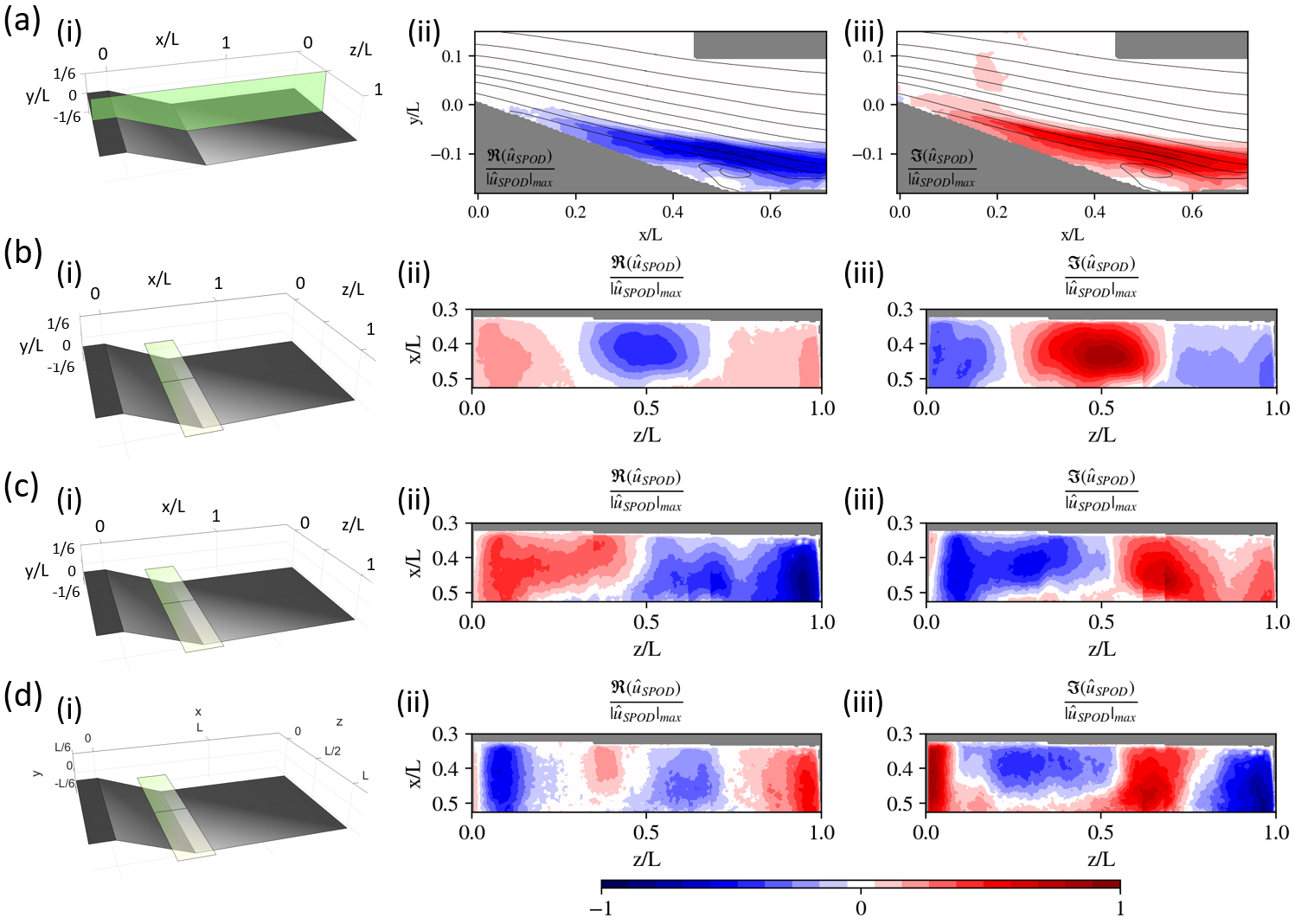}
    \caption{$u'$-component of leading symmetry-plane SPOD mode for $St=0.01$  (a), leading horizontal-plane SPOD mode in (b), subleading horizontal-plane SPOD mode (c), and third horizontal-plane SPOD mode. The transparent plane in the panel (a i) visualizes the location of the symmetry plane and in (b i), (c i) and (d i) the location of the horizontal plane. Panels (ii) show the normalized real part of the mode and (iii) show the normalized imaginary part of the mode. For normalization the maximum of the $u'$ mode magnitude is used. The dashed lines in (b), (c) and (d) highlight the position of the centerline. Coordinates are scaled with the spanwise domain length $L$.}
    \label{fig:SPOD_modes_combined}
\end{figure}

Figure \ref{fig:SPOD_modes_combined}(a ii and iii) shows the $u'$-component of the leading SPOD mode at $St=0.01$. This mode represents a large-scale streamwise elongated structure located above the separation bubble, extending from upstream of the separation point to well downstream of it. Qualitatively, the shape of this mode closely resembles the low-frequency modes identified by \cite{Cura2024} in the TSB generated over a flat plate.

Next, the horizontal-plane PIV dataset is analysed by means of SPOD to reveal the spanwise structure of the dominant flow dynamics. The corresponding spectra are shown in Figure \ref{fig:SPOD_spectrum_combined_unnormed}(b) and Figure \ref{fig:SPOD_spectrum_combined}(b). The results indicate that most of the energy is in the low frequency limit (at $St\approx0.01$). In contrast to the symmetry-plane spectrum, the horizontal-plane spectrum retains a non-negligible amount of energy in the two subleading modes in the frequency range between $St=0.01$ and $St=0.03$. The dynamics of the flow are thus dominated by three coherent structures rather than just one as observed in the streamwise-plane PIV.

The low-pass-filter behaviour that is observed in the symmetry plane SPOD is not visible in the horizontal-plane SPOD. We believe this is related to the location of the measurement plane. The horizontal PIV window cuts through the separation bubble $45$ mm above the bottom wall; the ramp is inclined at $\alpha=20^{\circ}$, and the separation point lies outside the window. 
The dependence of the measurement position on the observed low-pass-filter behaviour is also evident in the study of \citet{Cura2024}, who reported varying pressure power-spectral-density curves at different streamwise locations. 
Consequently, the horizontal-plane SPOD captures only part of the separation region. To resolve the characteristic low-pass behaviour, measurements would likely need to include a larger share of the shear layer formed by the TSB.

Figure~\ref{fig:SPOD_modes_combined}(b) shows the $u'$-component of the leading horizontal-plane SPOD mode at $St=0.01$. Note that the PIV domain spans the entire width between the walls. The centerline at $z/L=0.5$ is highlighted by a dashed line. A comparison of the real and imaginary components of the mode indicates that the wave pattern is stationary in the spanwise direction, characterizing the dynamics of this mode as a standing wave. The nodes, where the amplitude reaches its minimum, are located between the centerline and the walls at $z/L \approx 0.25$ and $0.75$. The antinodes, where the amplitude is maximum, are located at the centerline $z/L\approx 0.5$ and close to the walls $z/L \approx 0$ and $1$. Figure~\ref{fig:SPOD_modes_combined}(c) shows the $u'$-component of the subleading SPOD mode at $St=0.01$ in the horizontal plane. Similar to the leading mode, the subleading mode also exhibits characteristics of a standing wave. However, in this case, the node is located along the centerline (highlighted by a dashed line) at $z/L \approx 0.5$, while the antinodes are positioned close to the walls at $z/L \approx 0$ and $1$. Figure~\ref{fig:SPOD_modes_combined}(d) shows the $u'$-component of the third SPOD mode at $St = 0.01$. A standing-wave pattern is again observed, featuring three nodes at $z/L \approx 0.25$, $0.5$, and $0.75$.

In summary, the symmetry-plane SPOD analysis reveals low-frequency breathing dynamics between $St = 0.01$ and $St=0.05$, dominated by a single mode that spans the entire length of the separation bubble. For the horizontal-plane SPOD, three modes are dominant at $St \simeq 0.01$, all exhibiting standing wave behaviour with different spanwise node and antinode positions. The leading mode has two nodes at $z/L \approx 0.25$ and $0.75$, the subleading mode has one node along the centerline at $z \approx 0.5$, and the third mode has three nodes at $z/L \approx 0.25$, $0.50$ and $0.75$.Similar standing wave dynamics are observed for other frequencies in the low Strouhal regime, up to $St = 0.03$, but are not shown for brevity.
A connection between the two SPOD analyses is established by the fact that the leading streamwise SPOD mode corresponds to the leading horizontal SPOD mode, as both reach their maximum along the symmetry plane. In contrast, the subleading and the third horizontal-plane SPOD modes do not appear in the streamwise-plane SPOD analysis as their streamwise and crosswise velocity components are approximately zero along the centreline.

\subsection{Linear stability analysis of the 2D mean field} \label{sec:LSA_results}

A linear stability analysis (LSA) is performed on the mean flow in the symmetry plane. The analysis is conducted for spanwise wavenumbers in the range $\beta L_b / 2\pi = 0$ – $2$, using the perturbation ansatz given in Eq.~\ref{eq:ansatz_LSA} and the two-dimensional RANS mean field as base flow, as described in Section~\ref{sec:RANS}.

The eddy viscosity is prescribed as
\begin{align}
\nu_t = \gamma \, \nu_{t,\text{RANS}}(x,y),
\end{align}
where $\nu_{t,\text{RANS}}$ denotes the eddy viscosity obtained from the RANS solution, and $\gamma$ is a scaling factor. Using the unscaled RANS eddy viscosity ($\gamma = 1$) yields unstable eigenvalues. Performing a resolvent analysis on this mean flow would therefore require the discounted resolvent formulation \citep{Rolandi2024_invitation}, which we aimed to avoid.

\begin{figure}
\centering
\includegraphics[width=0.99\linewidth]{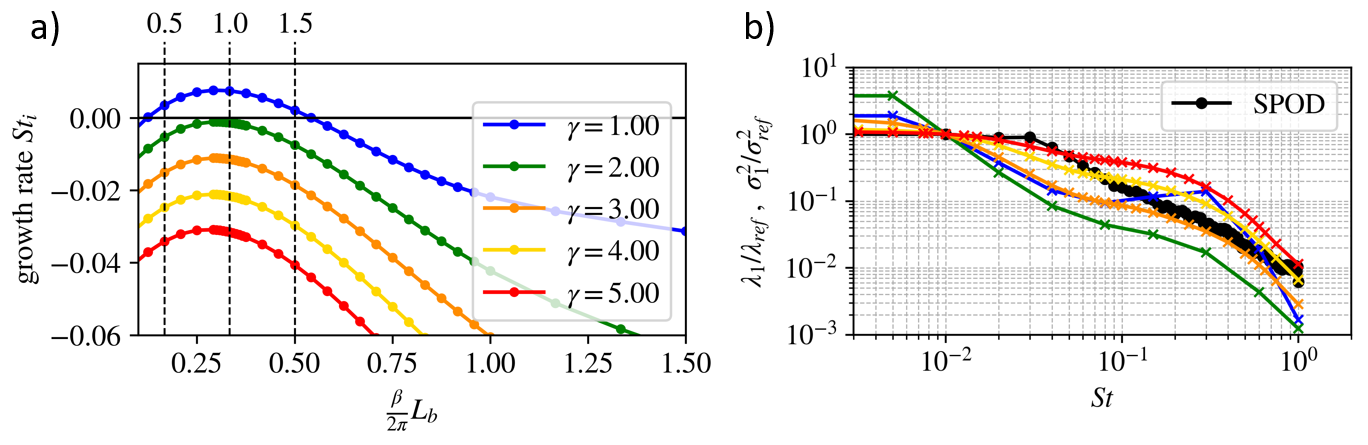}
\caption{(a) Stationary eigenvalues ($St_r = 0$) computed for five eddy-viscosity scaling factors, $\nu_t(x,y) = \gamma \, \nu_{t,\text{RANS}}(x,y)$ with $\gamma \in [1,2,3,4,5]$. (b) Leading resolvent gains for $\beta L_b / 2\pi = 1/3$ (equivalently $L/\lambda_z = 1$) compared with the SPOD spectrum in the symmetry plane for the same set of $\gamma$ values.}
\label{fig:gamma}
\end{figure}

Instead, we conducted a parametric study with different scaling factors to identify a suitable value of $\gamma$. Throughout the study, the maximum growth rates consistently occur at zero real frequency, $\omega_r = 0$, corresponding to stationary modes. Figure~\ref{fig:gamma}(a) shows the stationary eigenvalues ($\omega_r = 0$) obtained for varying $\gamma$. Increasing $\gamma$ reduces the growth rates of the stationary eigenmodes; for $\gamma \ge 2$, all eigenvalues are stable.
The choice of $\gamma$ can be further guided by considering its influence on the frequency response. \citet{Bugeat2022JfM} and \citet{SarrasPhD_thesis} demonstrate that the growth rate of a stationary eigenmode determines the cutoff frequency of the associated low-pass filter in the resolvent response. This trend is evident in Fig.~\ref{fig:gamma}(b), which shows the resolvent gain for $\beta L_b / 2\pi = 1/3$ ($L / \lambda_z = 1$), corresponding to the most energetic structure in the SPOD. As $\gamma$ increases beyond~2, the low-frequency plateau of the gain curve broadens, and the roll-off shifts to higher frequencies. This behaviour is characteristic of an increased low-pass filter cutoff frequency.
For the present analysis, $\gamma = 3$ is selected as an appropriate compromise, providing a realistic cutoff frequency of $St\approx 0.01$ while avoiding the numerical difficulties associated with excessively high eddy-viscosity values.
Note that a more rigorous approach to avoid the tuning of eddy viscosity would involve assimilating the RANS solution to the measured mean field and incorporating a linearized eddy-viscosity model into the linear operator \citep{SarrasJfM2024}. This, however, is beyond the scope of the present study. The close agreement between SPOD and linearized results justifies our approach.

\begin{figure}
\centering
\includegraphics[width=0.7\linewidth]{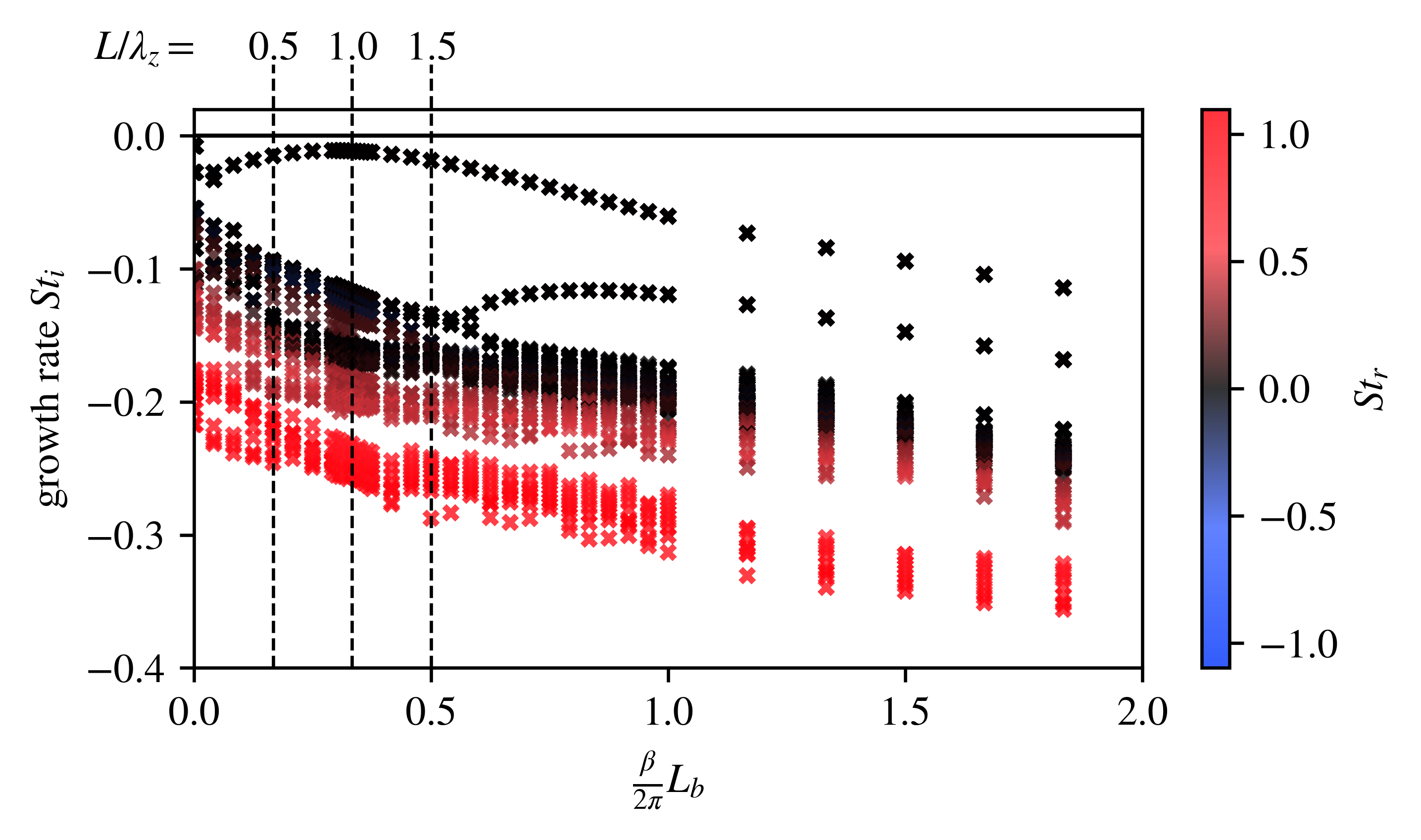}
\caption{Global linear stability eigenvalue spectra over a range of spanwise wavenumbers. The mode with the highest growth rate (for $\beta>0$) occurs at $\beta L_b / 2\pi = 0.292$ which corresponds to $L/\lambda_z=0.875$.}
\label{fig:LSA_spectrum}
\end{figure}

Figure~\ref{fig:LSA_spectrum} shows the computed eigenvalue spectra for $\gamma=3$ across the spanwise wavenumber range. The maximum growth rates consistently occur at zero real frequency, $\omega_r = 0$, corresponding to stationary modes. 
The wavenumber corresponding to the maximum growth rate is close to $L/\lambda_z=1$, indicating that the most amplified spanwise wavelength is approximately equal to the channel width $L$.

\begin{figure}
\centering
\includegraphics[width=0.8\textwidth]{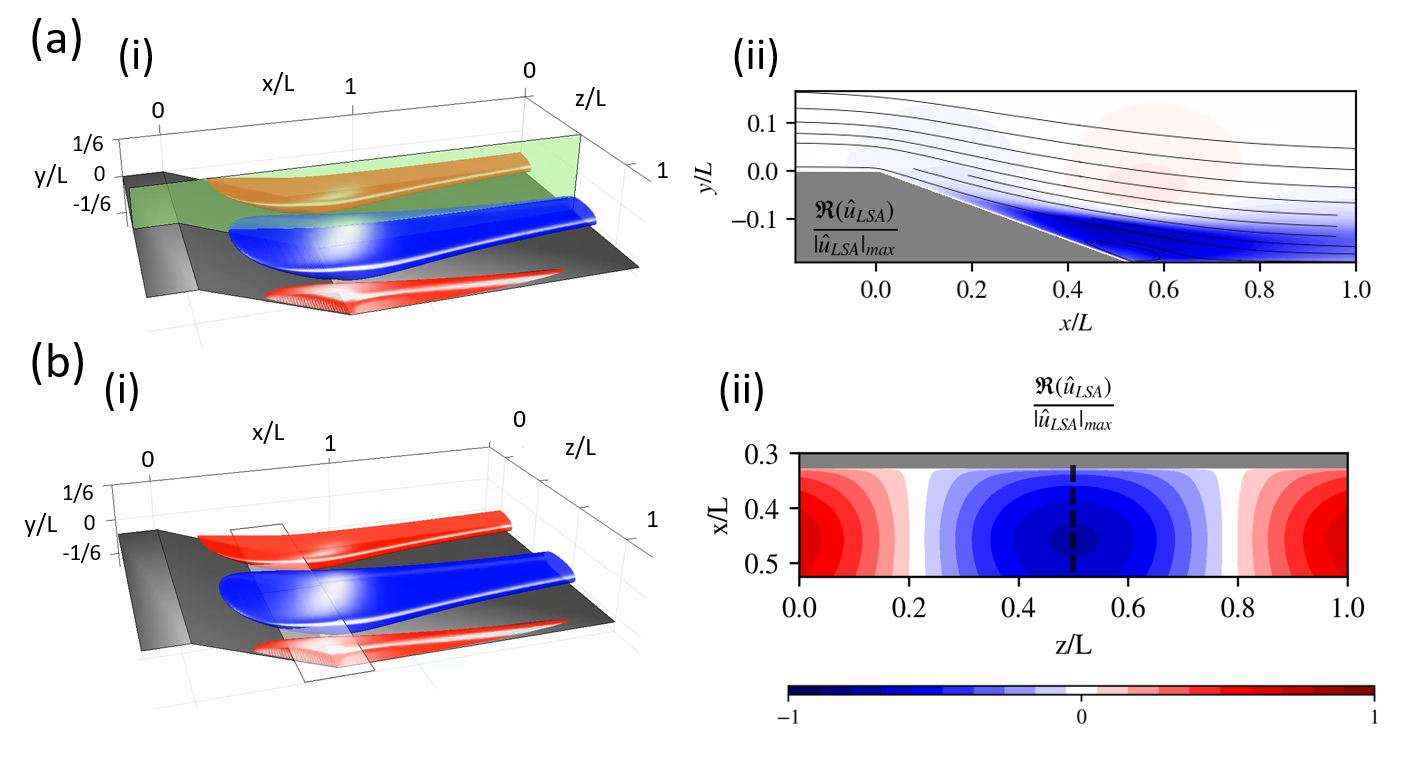}
\caption{$u$-component of stationary eigenmode ($St = 0$) for $\beta \frac{L_b}{2\pi}=0.292$  which corresponds to $L/\lambda_z=0.875$. This eigenmode corresponds to the eigenvalue with the maximum growth rate. Panel (a) shows the mode in the and Panel (b) the standing wave model using the eigenmode in the horizontal plane. The panels (i) show 3D isosurfaces of the real part of the mode. The transparent plane in the panel (a i) visualises the location of the symmetry plane and in (b i) the location of the horizontal plane. Panels (ii) show the real part of the mode. The dashed lines in (b) highlight the position of the centerline. Coordinates are scaled with the spanwise domain length $L$. It is noted that at zero frequency only the real part contributes to the mode.}
\label{fig:LSA_mode}
\end{figure}

The eigenmode with maximum growth rate is shown in Figure~\ref{fig:LSA_mode}. It corresponds to a stationary mode ($\omega_r = 0$) at $\beta \frac{L_b}{2\pi}=0.292$ and is positioned above the separation bubble, extending from the separation region to well downstream. The mode shape closely resembles the leading SPOD mode obtained in the symmetry plane (see Fig.~\ref{fig:RA_modes}(a ii and iii)), indicating that the breathing motion is likely associated with a stationary modal mechanism whose growth rate depends strongly on the spanwise wavenumber. These findings are consistent with recent studies by \citet{Cura2024, Cura2025, SavarinoJFM2025}.

In summary, the LSA reveals a stationary eigenmode whose spatial structure closely matches the observed patterns and thus provides a plausible modal origin for the low-frequency dynamics. To assess how this mode imprints on non-zero frequencies, we next consider resolvent analysis.

\subsection{Resolvent analysis of the 2D mean field}
Resolvent analysis is performed on the mean field in the symmetry plane. The analysis is conducted for a spanwise wavenumbers range between $\beta  \frac{L_b}{2\pi} = 0$ and $\beta  \frac{L_b}{2\pi} = 2$ and frequency range between $St = 0$ and $St = 1$. We use the ansatz described in Eq.~\ref{eq:travelling_wave}, the two-dimensional RANS mean field as base flow, as described in Section~\ref{sec:RANS}, and an eddy viscosity scaling factor $\gamma=3$, as described in Section~\ref{sec:LSA_results}. For each combination of $St$ and $\beta$, the resolvent analysis provides a gain value $\sigma$ and the corresponding forcing and response modes. It is important to note that, since the base flow is assumed to be homogeneous in spanwise direction, this ansatz does not take into account the boundary conditions at the sidewalls.

\begin{figure}
    \centering 
    \includegraphics[width=1.0\linewidth]{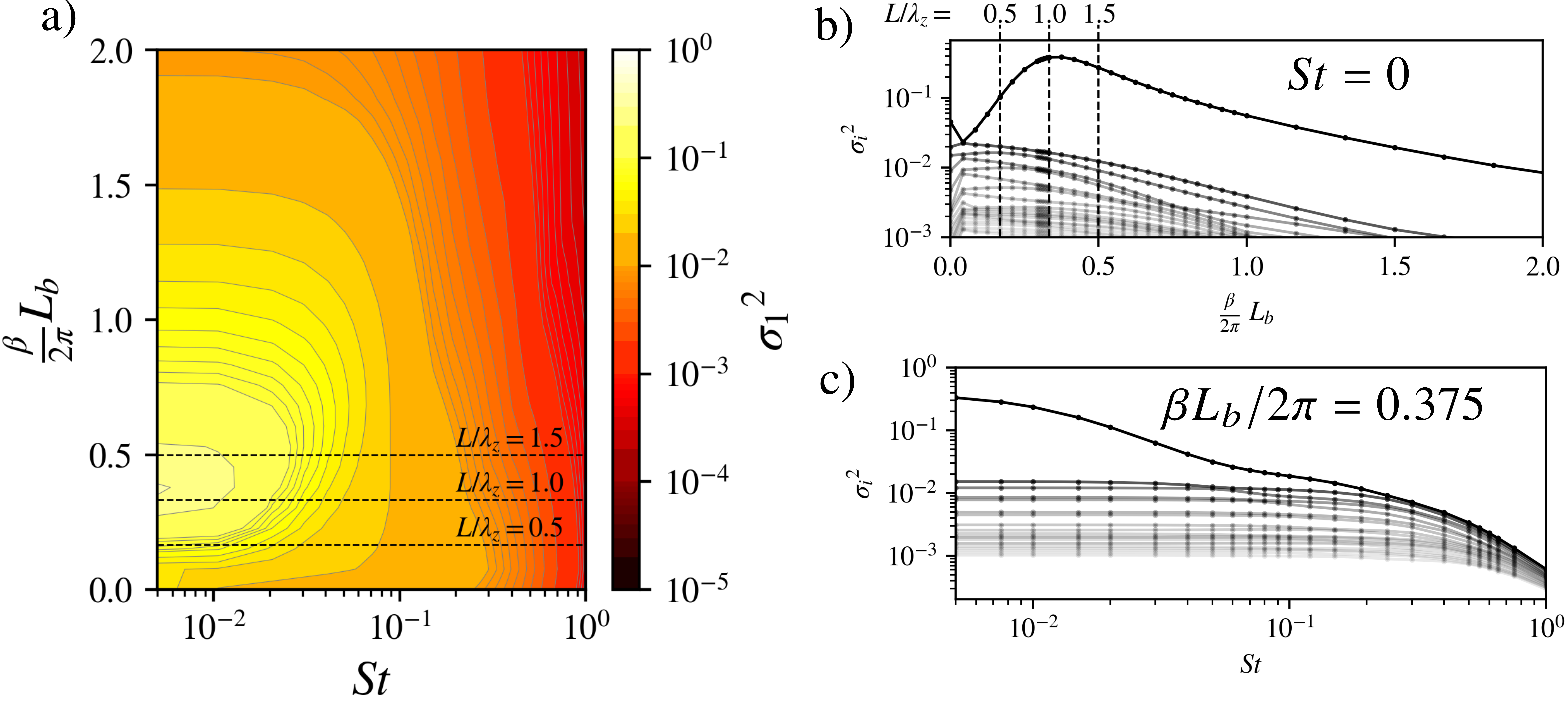}
    \caption{Gain values of the leading mode for various frequencies ($St$) and nondimensionalised spanwise wavenumbers ($\beta L_b / 2\pi$) from resolvent analyses (a). Panel (b) shows the gain of the first 20 resolvent modes at $St = 0$, where the maximum occurs at $\beta L_b / 2\pi = 0.375$. Panel (c) presents the gains of the first 20 modes at this peak spanwise wavenumber for different frequencies. Dashed lines in (a) and (b) indicate the wavenumbers for which the spanwise wavelength $\lambda_z$ equals 0.5, 1, and 1.5 times the spanwise domain width $L$.}
    \label{fig:RA_gains}
\end{figure}

The distribution of all gain values for the leading mode is illustrated in Fig. \ref{fig:RA_gains}.
The horizontal axis represents the non-dimensional frequency $St$, while the vertical axis shows the non-dimensional spanwise wavenumber $\beta L_b / 2\pi$. A value of $\beta L_b / 2\pi = 1/3$ corresponds to a spanwise wavelength $\lambda_z$ equal to the channel width $L$, signifying one full period across the span. Note that in numerical simulations with periodic boundary conditions, only integer values of $L/\lambda_z$ would be permissible, limiting the range of spanwise wavelengths that can be resolved.

The results in Figure \ref{fig:RA_gains} show that the maximum gain occurs near $\beta L_b / 2\pi = 0.375$ for frequencies below $St = 0.02$. At higher frequencies, the gain values drop rapidly.

Figure \ref{fig:RA_gains}(b) and (c) presents the gain values of the first 20 resolvent modes at the spanwise wavenumber with the highest gain $\beta L_b / 2\pi = 0.375$, and at $St = 0$. The gain of the leading mode is several orders of magnitude higher than that of all other modes. These results indicate that, based on the resolvent analysis, the flow dynamics can be effectively described as low-rank.

\begin{figure}
\centering
\includegraphics[width=0.6\linewidth]{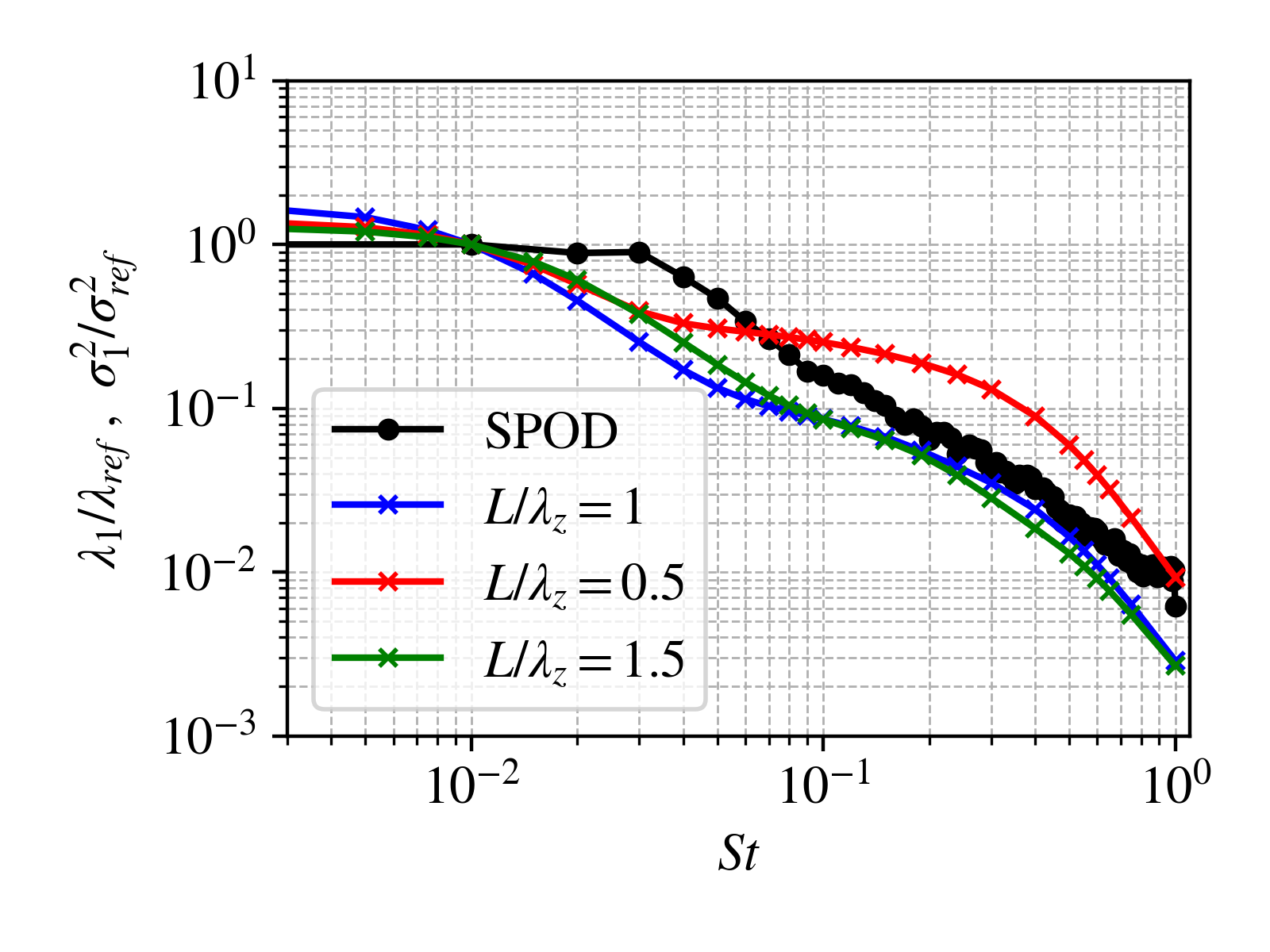}
\caption{Resolvent gains of the leading mode for the dominant spanwise wavenumbers, $\beta L_b / 2\pi = 1/6$ ($L/\lambda_z = 0.5$), $\beta L_b / 2\pi = 1/3$ ($L/\lambda_z = 1$), and $\beta L_b / 2\pi = 1/2$ ($L/\lambda_z = 1.5$). The eigenvalues of the leading symmetry plane SPOD modes are shown for comparison. All quantities are normalised by their respective values at the reference frequency $St_{ref} = 0.01$ to enable qualitative comparison.}
\label{fig:SPOD_RA_gain_comparison}
\end{figure}

For a comparison between the data-driven SPOD results and the physics-based resolvent model, Fig.\ref{fig:SPOD_RA_gain_comparison} shows the leading symmetry-plane SPOD eigenvalues and the leading resolvent gains for the spanwise wavenumber $\beta L_b / 2\pi = 1/6$ ($L/\lambda_z = 0.5$), $\beta L_b / 2\pi = 1/3$ ($L/\lambda_z = 1$), and $\beta L_b / 2\pi = 1/2$ ($L/\lambda_z = 1.5$). This corresponds to the wavenumbers of the three most dominant modes observed in the horizontal plane. $L/\lambda_z = 1$ is close to the wavenumber of maximum growth rate obtained from the global stability analysis, see Fig.~\ref{fig:LSA_spectrum}. Note that the resolvent gains and the SPOD eigenvalues are not directly comparable: the resolvent is evaluated at discrete spanwise wavenumbers, whereas the symmetry-plane SPOD is not strictly restricted to a single spanwise wavenumber and may include contributions from the entire spanwise spectrum. All quantities are normalised by their respective values at the reference frequency $St_{ref} = 0.01$ to enable qualitative comparison.

The low-pass-filter trend of the SPOD eigenvalues is reproduced by the resolvent gain; however, the cutoff frequency is underpredicted. This discrepancy is expected given the following factors. First, the resolvent model predicts low-rank dynamics over a broad range of spanwise wavenumbers, implying that the leading symmetry-plane SPOD eigenvalue spectrum results from contributions of multiple modes and therefore cannot be matched by a single resolvent mode at a single wavenumber; frequency-dependent contributions from several wavenumbers are expected. Second, SPOD and resolvent analyses are theoretically identical only under white-noise forcing~\citep{Towne2018}. Since the true forcing statistics are unknown, some deviation between the two spectra is expected. Third, as shown in Fig.~\ref{fig:gamma}(b), the cutoff frequency is influenced by the eddy viscosity scaling $\gamma$. Given these limitations, the level of agreement between SPOD eigenvalues and resolvent gain is considered satisfactory.

The findings presented in Fig.~\ref{fig:RA_gains} share similarities with previous studies, particularly with the work of \cite{Cura2024}. Their results also show maximum amplification at low frequencies and nonzero spanwise wavenumbers, along with a secondary peak at higher frequencies (around $St = 1$ to $2 \times 10^{-1}$), which they attribute to shear-layer modes. However, an important distinction is that \cite{Cura2024} considered a different base flow: a pressure-induced turbulent separation bubble over a flat plate. The spanwise wavenumber differs slightly between the two cases: when normalised by the average bubble length $L_b$, the spanwise wavenumber at which they find the maximum resolvent gain is about 20 percent larger than the corresponding value in our setup. Additionally, they identify a weakly damped stationary global eigenmode at approximately one half of the wavenumber associated with the maximum resolvent gain in our case.

A similar behaviour, comparable to the findings of \cite{Cura2024}, was also reported by \cite{Hao2023}, who studied shock wave–turbulent boundary layer interactions over a compression corner using global stability analysis and resolvent analysis. His analysis revealed an unstable stationary mode at nonzero spanwise wavenumbers, with the spanwise wavelength scaling consistently with the separation bubble length $L_b$ across different ramp angles and Reynolds numbers. Furthermore, in his resolvent analysis, he also found maximum gain at nonzero spanwise wavenumbers for a frequency $St=0.015$, which closely matches the relevant frequency range observed in our study. However, it is important to note that his configuration differs significantly from ours, as he examines a high-Mach-number flow at Reynolds numbers approximately 60-130 times higher than those in our study. Despite these differences, his findings suggest a possible connection between low-frequency dynamics and the spanwise wavenumber of the separation bubble in relation to the streamwise separation length, which appears to be a robust feature across different flow regimes.

While the studies by \cite{Cura2024} and \cite{Hao2023} show clear amplification at nonzero spanwise wavenumbers, the study by \cite{Wu2019} does not exhibit dominant modes with similar spanwise characteristics at comparable frequencies. In the spanwise direction, their computational domain is narrower than the length of a separation bubble, $L_b$, whereas in our case, the spanwise wavelength with the highest gain is approximately equal to the bubble length ($\lambda_z / L_b \approx 1$). This suggests that the phenomena observed in our study could not be resolved within the domain constraints of \cite{Wu2019}.  
The results of the resolvent analysis show that there is a linear amplification mechanism intrinsic to the 2D mean flow that occurs within certain ranges of spanwise wavenumbers and low frequencies. This dependence on the spanwise wavenumber aligns with previous observations, such as those by \citet{Cura2024}.
Similar behaviour is reported by \citet{SarrasJfM2024}, who conducted global linear stability analysis on a RANS mean flow. Their findings reveal stationary three-dimensional global modes becoming unstable only within certain ranges of angles of attack and particular spanwise wavenumbers, a feature consistent with the wavenumber selectivity identified in the present resolvent analysis. It is important to note that the gain derived from the 2D mean flow does not take into account the boundary conditions at the walls. Recent results by \citet{Cura2025} indicate that variations in reverse flow and separation-bubble length modify the growth rates, whereas the dominant spanwise wavenumbers are much less sensitive. This may explain why experimental and modelling results show good agreement despite differences in reverse flow and separation-bubble length, as discussed in Section~\ref{sec:RANS}.

\subsection{Resolvent-based 3D standing wave model}

In the following, we construct a standing wave model based on the resolvent analysis, as outlined in Eq. \ref{eq:standing_wave}. While resolvent analysis allows for arbitrary spanwise wavenumbers $\beta$, the reflective slip-wall boundary condition is only satisfied for discrete values of $L/\lambda_z=0.5, 1, 1.5, \dots$ which corresponds to $\beta \frac{L_b}{2\pi} = 1/6,\ 1/3,\ 1/2$, as discussed in Chapter \ref{sec:standing_wave_model}.

\begin{figure}
    \centering 
    \includegraphics[width=1.0\textwidth]{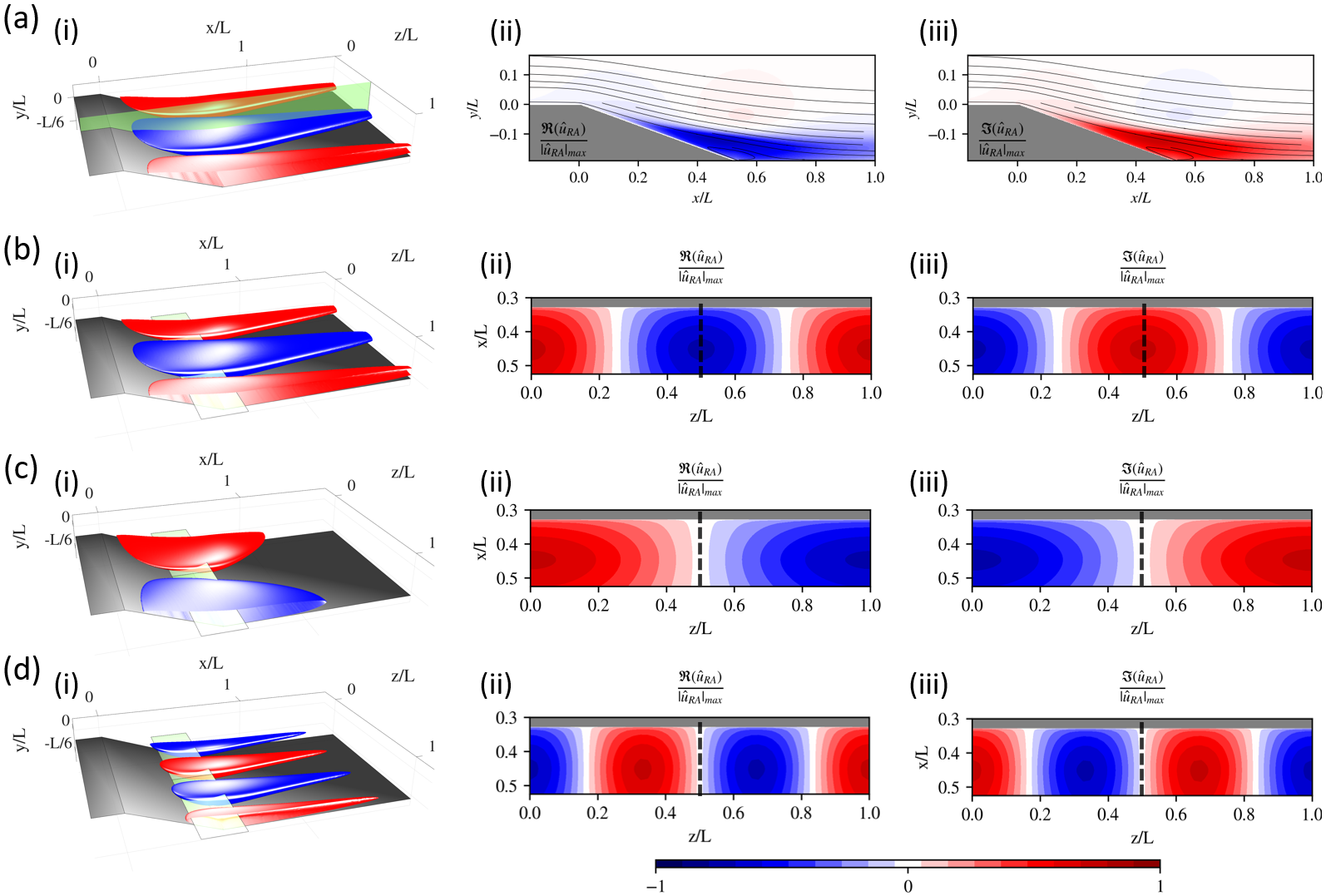}
    \caption{Standing wave model ($u$-component) for $St=0.01$ and $\beta \frac{L_b}{2\pi}=1/3$ ($L/\lambda_z=1$) in symmetry plane (a) and in horizontal plane  (b), for $\beta \frac{L_b}{2\pi}=1/6$ ($L/\lambda_z=0.5$) in horizontal plane (c), and for $\beta \frac{L_b}{2\pi}=1/2$ ($L/\lambda_z=1.5$) in horizontal plane (d). The panels (i) show 3D isosurfaces of the real part of the mode. The transparent plane in the panel (a i) visualises the location of the symmetry plane and in (b i), (c i), and (d i) the location of the horizontal plane. Panels (ii) show the real part of the mode and (iii) show the imaginary part of the mode. For normalization the maximum of the $u$ mode magnitude is used. The dashed lines in (b), (c), and (d) highlight the position of the centerline. Coordinates are scaled with the spanwise domain length $L$.}
    \label{fig:RA_modes}
\end{figure}

Figure \ref{fig:RA_modes} presents the standing wave model for three selected spanwise wavenumbers: $\beta \frac{L_b}{2\pi}=1/6$ ($L/\lambda_z=0.5$), $\beta \frac{L_b}{2\pi}=1/3$ ($L/\lambda_z=1$) and $\beta \frac{L_b}{2\pi}=1/2$ ($L/\lambda_z=1.5$). These values correspond to the lowest wavenumbers that match the reflective slip boundary conditions.
It is worth noting that the spanwise wavenumbers selected for the standing wave model do not coincide with the peak gain predicted by the traveling-wave resolvent analysis, which occurs near $\beta _bL / 2\pi = 0.375$. Although the gain at the selected wavenumbers remains high, it is not maximal. This suggests that individual gain values may be less relevant compared to the ability of a mode to satisfy the reflective boundary conditions imposed by the domain. Indeed, the dominant structures observed in the experiment correspond to the longest spanwise wavelengths that are compatible with the sidewalls. 

A comparison between the standing wave model and SPOD results in the symmetry plane reveals that the model accurately captures the shape of the dominant mode. The mode extends from upstream of the separation bubble to well beyond the reattachment region. The 3D visualization (Fig. \ref{fig:RA_modes}(i)) further illustrates that the mode forms a streamwise-elongated structure. Additionally, the standing wave model successfully captures mode dynamics in the horizontal plane for both the leading mode (compare Figs. \ref{fig:SPOD_modes_combined}(b) and \ref{fig:RA_modes}(b) for $\beta \frac{L_b}{2\pi}=1/3$ ($L/\lambda_z=1$)) and the subleading mode (compare Figs. \ref{fig:SPOD_modes_combined}(c) and \ref{fig:RA_modes}(c) for $\beta \frac{L_b}{2\pi}=1/6$ ($L/\lambda_z=0.5$)). In both cases, the model accurately predicts the positions of nodes and antinodes. Further comparison with Fig. \ref{fig:waves}(c ii) confirms that these modes correspond to the two largest possible standing waves that fit within the channel.

\begin{figure}
    \centering 
    \includegraphics[width=0.8\textwidth]{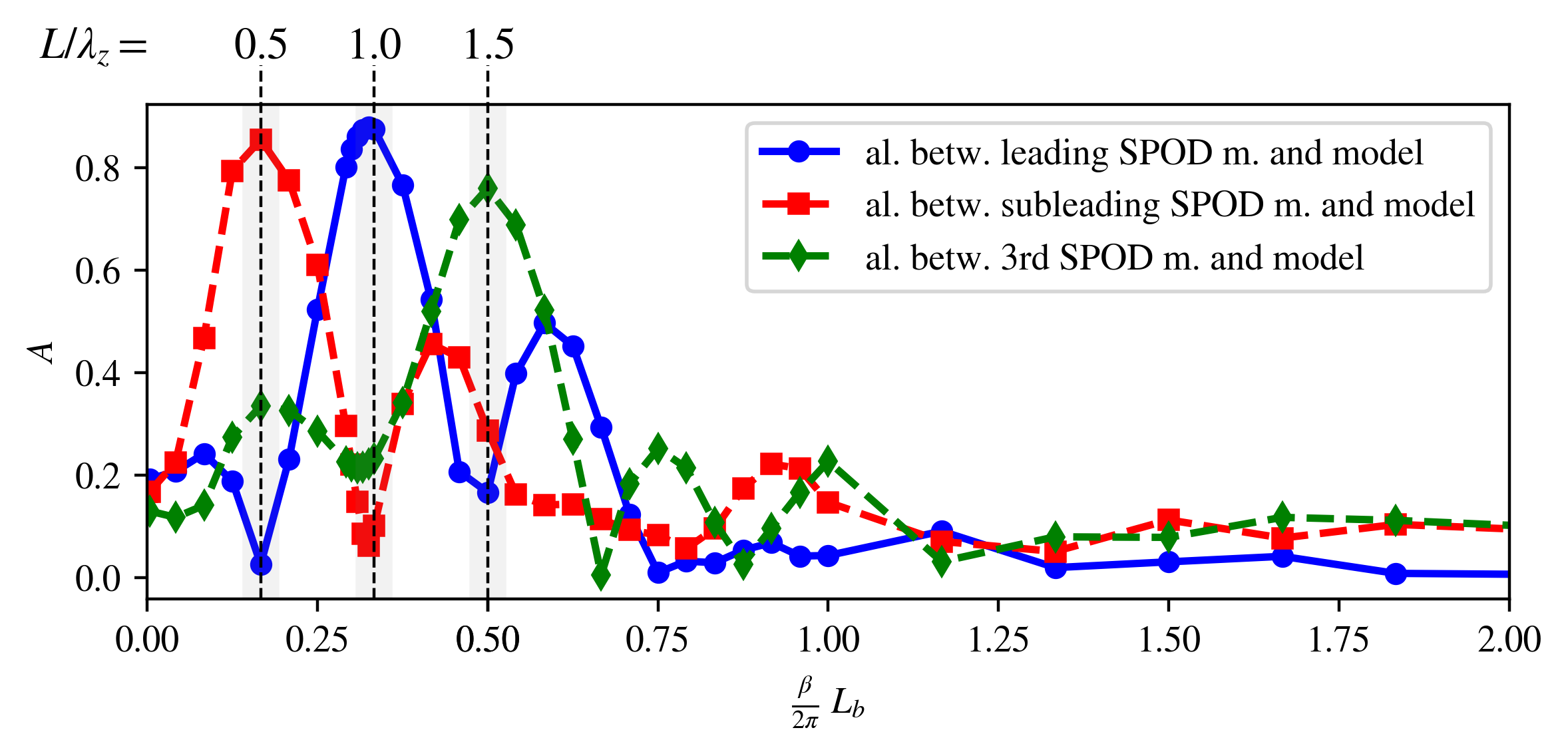}
    \caption{Alignment between SPOD mode in horizontal plane and resolvent models for $St=$0.01. The blue, red and green lines show the alignment between the standing wave model and the leading, subleading, and second subleading, SPOD mode, respectively.}
    \label{fig:alignment_horizontalplane}
\end{figure}

The alignment between the SPOD modes and the resolvent-based standing wave model is shown in Figure \ref{fig:alignment_horizontalplane} for a fixed frequency of $St=0.01$.

The alignment is given by 
\begin{equation}
A = \frac{|\langle \hat{u}_{SPOD},\hat{u}_{RA} \rangle_{L2}|}{\sqrt{\langle \hat{u}_{SPOD},\hat{u}_{RA} \rangle_{L2} \ \langle \hat{u}_{SPOD},\hat{u}_{RA} \rangle_{L2}}} \, 
\end{equation}
where $\langle \cdot \rangle_{L2}$ denotes the discrete inner product, given by 
\begin{equation}
    \langle \cdot \rangle_{L2} = \sum_{m,n} A(x_n, y_m)^\star B(x_n, y_m) \ ,
\end{equation}
with $(\cdot)^\star$ denoting the complex conjugate.

The alignment is a commonly used measure to compare SPOD and resolvent modes which provides a phase-independent shape comparison, ensuring that only the structural similarity of the fields is considered, independent of magnitude \cite{Cavalieri2013, Pickering2021, vonSaldern_JFM_2024}. The results indicate that the highest alignment between the leading SPOD mode and the standing wave model occurs at $L/\lambda_z=1$, while for the subleading SPOD mode, the best agreement is observed at $L/\lambda_z=0.5$, and for the third SPOD mode it is observed at $L/\lambda_z=1.5$. The results in Fig.~\ref{fig:alignment_horizontalplane} are shown for $St=0.01$ only; however, the standing-wave model remains a good approximation at $L/\lambda_z=1$, consistent with the leading SPOD mode up to $St \approx 0.03$ (not shown). 
Notably, the alignment plots confirm a posteriori that $L$ is the appropriate length scale for the standing wave model. This finding is particularly remarkable given that the 3D mean flow is not strictly homogeneous in the spanwise direction, as it exhibits a U-shaped mean separation line, a characteristic feature recently demonstrated by \citep{SteinfurthPINN_PoF2024} and shown in Fig. \ref{fig:3D_meanflow}(a).

The standing wave model exhibits good agreement with the experimental SPOD modes in both the symmetry and horizontal planes. Beyond this validation, the model can be extended to construct a fictitious total flow field by superimposing the standing wave modes onto the mean flow. This approach provides an intuitive way to explore the spatial and temporal structure of the three-dimensional dynamics of the low frequency breathing motion.

\begin{figure}
\centering
\includegraphics[width=1.0\columnwidth]{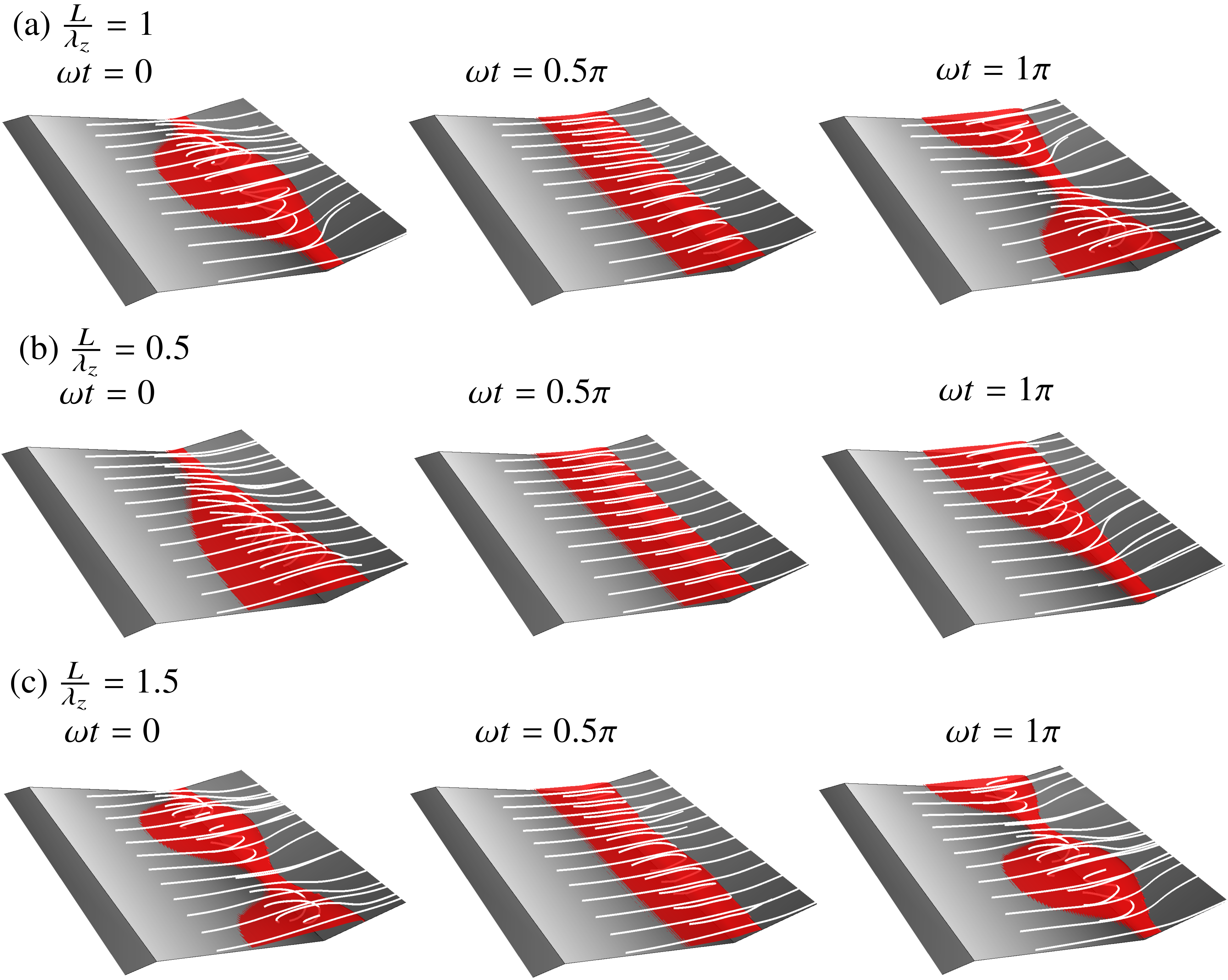}
\caption{Snapshots of a three dimensional total flow simulation for $St=$ 0.01 using the mean flow with the scaled standing wave model according to Eq.~\ref{eq:3DSimulation}. The blue semitransparent iso-surface shows the separation bubble with $u = 0$. The black lines show streamlines anchored upstream and downstream of the separation point. The motion is periodic with a full cycle described by $\omega t= 2\pi$. Due to the standing wave nature, the second half of the cycle mirrors the first in reverse order. Panel (a) shows the mode for $\beta \frac{L_b}{2\pi}=1/3$ ($L/\lambda_z=1$) which is the mode that shows closest similarity with the leading SPOD mode in the experimental data. Panel (b) shows the mode for $\beta \frac{L_b}{2\pi}=1/6$ ($L/\lambda_z=0.5$) and panel (c) for $\beta \frac{L_b}{2\pi}=1/2$ ($L/\lambda_z=1.5$) which show closest similarity with the second and third SPOD mode in the experimental data.}
\label{fig:bubble_animation}
\end{figure}

Figure \ref{fig:bubble_animation} presents a three-dimensional visualization of a fictitious total flow field, constructed as

\begin{align} 
\boldsymbol{q}(x,y,z,t) = \boldsymbol{\bar{q}}(x,y)+\alpha \ \boldsymbol{q}'_{\text{sw}}(x,y,z,t) 
\label{eq:3DSimulation}
\end{align} 

where $\boldsymbol{\bar{q}}(x,y)$ denotes the two-dimensional mean flow in the symmetry plane and $\boldsymbol{q}'_{\text{sw}}(x,y,z,t)$ denotes the standing wave mode given by Eq.~\ref{eq:standing_wave} that is scaled by an arbitrary factor $\alpha$.

To ensure a consistent reference magnitude, the standing wave mode $\boldsymbol{q}'_{\text{sw}}$ has been normalised by its maximum absolute value, such that $||\boldsymbol{\hat{q}_{sw}} ||=1$. The scaling factor $\alpha$ is deliberately set to a relatively high value ($\alpha= \frac{2}{3} u_{\infty}$) to enhance the illustrative effect of the figure. While this results in a certain degree of over-amplification, it makes the oscillatory dynamics more clearly visible.

Figure \ref{fig:bubble_animation}(a) depicts snapshots of the first half period of the total flow animation using the standing wave model with $\beta \frac{L_b}{2\pi}=1/3$ ($L/\lambda_z=1$). This is the spanwise wavenumber that matches the leading SPOD mode best. Here, the separation bubble undergoes a breathing-like oscillation: when the bubble is constricted in the symmetry plane, it expands near the sidewalls, and vice versa. This is consistent with dedicated measurements by \citep{Steinfurth_XLASER_2022}. The streamlines reveal a spanwise velocity component directed towards the regions of local bubble enlargement, supporting the dynamic redistribution of the separation region. The enlargement of the bubble oscillates from the centre outwards and back.  

Figure \ref{fig:bubble_animation}(b) illustrates the total flow animation with $\beta \frac{L_b}{2\pi}=1/6$ ($L/\lambda_z=0.5$), which closely matches the subleading SPOD mode. This mode exhibits a meandering motion of the separation bubble, shifting from one sidewall to the other. The enlargement of the bubble oscillates between the two sidewalls. Figure \ref{fig:bubble_animation}(c) illustrates the total flow animation with $\beta \frac{L_b}{2\pi}=1/2$ ($L/\lambda_z=1.5$), which closely matches the third SPOD mode. This mode exhibits two bubble enlargements oscillating between the sidewalls.

In summary, the results demonstrate that the resolvent-based standing wave model effectively captures the dominant low-frequency modes in the flow. The model successfully reproduces the spatial structure of both the leading and subleading SPOD modes, exhibiting strong agreement in both the symmetry and horizontal planes. Furthermore, the model, constructed as a superposition of two opposing resolvent waves with a slip-wall condition, accurately represents the standing wave behaviour and mode shape in the spanwise direction that are observed in the experimental data. The alignment results confirm {\itshape a posteriori} that spanwise domain width $L$  is the appropriate length scale for the standing wave model, despite the mean flow not being strictly homogeneous in the spanwise direction. Additionally, the standing wave model enables the construction of an illustrative representation of the three-dimensional motion of the separation bubble.

Remarkably, the standing-wave resolvent model based on a 2D mean flow captures the low-frequency spanwise dynamics of the TSB, although the average TSB flow is strongly three-dimensional, as seen in Fig.~2. This suggests that low-frequency dynamics in turbulent separated flows are not necessarily driven by their three-dimensional character.

\subsection{Physical Mechanism Driving Low-frequency Breathing}
    
In recent years, the physical mechanism underlying the breathing dynamics has been widely discussed, and several explanations have been proposed. These include Görtler-type instability, (low spanwise wavenumber) centrifugal instability, the lift-up mechanism, and combinations thereof. In the following, we summarise the key studies that proposed these mechanisms and assess their relevance to our configuration. 

\subsubsection{Görtler-type Instability}

Investigating a TSB on a flat plate, \cite{Wu2019} suggest that the low-frequency mode is associated with Görtler vortices. These vortices arise due to a centrifugal instability induced by curved streamlines and the associated centrifugal forces. The most amplified wavelengths of Görtler vortices depend on boundary-layer thickness, velocity, and surface curvature \citep{Goertler1954}. Typically, the wavelength is of the order of the local boundary-layer thickness, between $\lambda \approx \delta$ and $\lambda \approx 2\delta$ \citep{SmitsDussauge2006}. 
\cite{Wu2019} applied this approach to a turbulent mean field by incorporating an effective viscosity combining molecular and eddy viscosity, and computed the characteristic wavelength. Their analysis reveals that the Görtler instability threshold is exceeded for counter-rotating streamwise vortices with spanwise wavelengths about one-fifteenth of the mean separation-bubble length. 
\cite{Cura2024} applied the same methodology and found that the observed waves were significantly larger than the local boundary-layer scale. Consequently, they suggested that these structures are unlikely to be related to Görtler vortices. The spanwise wavelength of the dominant mode was approximately 6.3 times larger than the mean separation-bubble length.

In our setup, the spanwise wavelength that corresponds to the leading SPOD mode is approximately three times larger than the mean separation-bubble length, and six times larger for the case corresponding to the subleading SPOD mode.
Further, by comparing the vertical bubble height with the average separation length in Figure~\ref{fig:RANS_mean_field}, the maximum boundary-layer thickness can be roughly estimated as $\delta_{99}(x) \approx 0.3 L_{b}$. This implies that the spanwise wavelengths of the dominant coherent structures are 10–20 times larger than the maximum local boundary-layer thickness, which makes them highly unlikely to be driven by a Görtler-type instability. 

\subsubsection{Centrifugal-type Instability}

Centrifugal instability is not restricted to the Görtler type; several curvature-driven variants may occur. In laminar separation bubbles, \cite{Barkley2002, Rodrguez2013} and \citet{Gallaire2007} identified centrifugal mechanisms, and more recently, \citet{SavarinoJFM2025} reported similar effects in a transitional separation bubble.
Such instabilities are modal in nature and have been associated with stationary eigenmodes \citep{Gallaire2007, Rodrguez2013, SavarinoJFM2025}. Since the present low-frequency breathing originates in an eigenmode, a centrifugal mechanism represents a plausible explanation.
To assess its likelihood, we evaluate the Rayleigh discriminant following the approach of \citet{Bayly1988}, \citet{Sipp_Jacquin_PoF_2000}, \cite{Barkley2002}, \citet{Gallaire2007}, and \citet{SavarinoJFM2025}.

In the formulation used by \citet{Gallaire2007}, the flow is centrifugally unstable if there exists a streamline~$\psi$ such that, for every point~$x_s$ on the streamline, the Rayleigh discriminant $R$ is negative, which is defined as
\begin{equation}
R = \frac{2 |\mathbf{U}| \Omega}{r},
\end{equation}
where $|\mathbf{U}|=\sqrt{\overline{u}^2+\overline{v}^2}$ is the local velocity magnitude, $\Omega=\frac{d\overline{v}}{dx}-\frac{d\overline{u}}{dy}$ the vorticity component in the x-y plane, and $r$ the local algebraic radius of curvature, defined by
\begin{equation}
r = \frac{|\mathbf{U}|^3}{(\nabla \psi) \cdot [\mathbf{U} \cdot \nabla\mathbf{U}]}\ .
\end{equation}
To evaluate $r$ at any point $(x,y)$, in the 2D meanflow ($\overline{w}=0$) we use
\begin{equation}
r(x, y) = \frac{|\mathbf{U}|^3}{\overline{u} \, a_y - \overline{v} \, a_x}  \,
\end{equation}
with 
\begin{equation}
a_x = \overline{u} \, \frac{\partial \overline{u}}{\partial x} + \overline{v} \, \frac{\partial \overline{u}}{\partial y},  \ \ 
a_y = \overline{u} \, \frac{\partial \overline{v}}{\partial x} + \overline{v} \, \frac{\partial \overline{v}}{\partial y} \ .
\end{equation}

The criterion is satisfied only if $R<0$ along the entire (closed) streamline, which, by definition, can occur only within the recirculation bubble. In the work of \citet{Gallaire2007}, no closed streamline was found to exhibit $R<0$ everywhere, but regions of negative $R$ were identified inside the bubble. They argued that fluid particles traveling along these streamlines periodically pass through regions with $R<0$ and are therefore locally susceptible to centrifugal instability.

\begin{figure}
\centering
\includegraphics[width=0.8\textwidth]{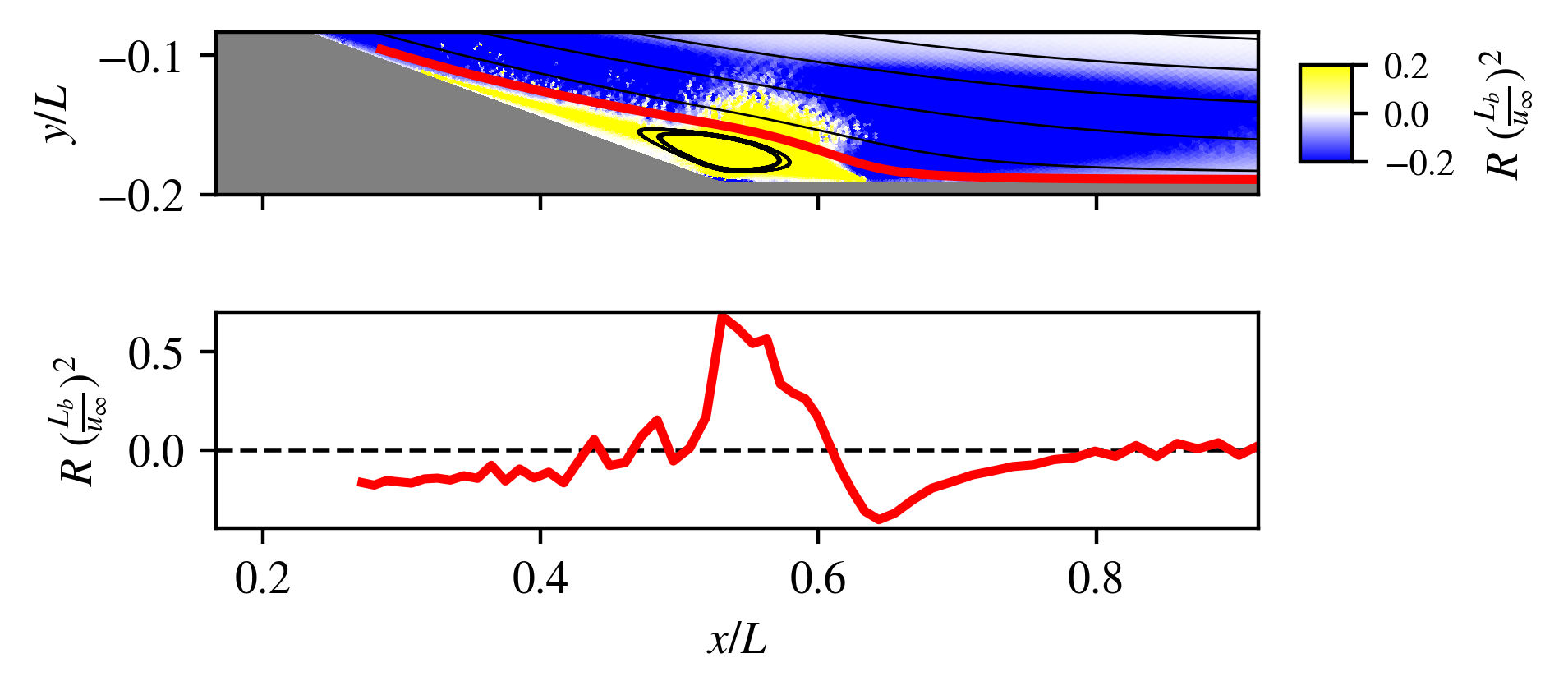}
\caption{Rayleigh discriminant evaluated on the mean flow. The red line in the upper panel indicates the streamline along which the Rayleigh discriminant $R$ is plotted in the lower panel. Coordinates are scaled with the spanwise domain length $L$.}
\label{fig:Rayleigh_Criterion}
\end{figure}

Figure~\ref{fig:Rayleigh_Criterion} shows the spatial distribution of the Rayleigh discriminant for the present flow. Negative values occur predominantly outside the separation bubble, while the interior of the bubble remains positive throughout. Although no closed streamline exhibits $R<0$ along its entire length, regions of negative $R$ appear near the separation and reattachment points. The lower panel of Fig.~\ref{fig:Rayleigh_Criterion} displays $R$ along a streamline closely following the bubble boundary, where $R$ becomes negative upstream and downstream of the recirculation zone. Hence, the formal instability criterion is not satisfied, but the presence of localised regions with $R<0$ suggests a susceptibility to centrifugal effects in these areas. A similar argument is followed by~\cite{SavarinoJFM2025} for their transitional flat plat bubble. Taken together with the presence of a weakly damped zero-frequency global eigenmode, and following the arguments of \citet{Gallaire2007} and \citet{SavarinoJFM2025}, it is plausible that centrifugal effects contribute to the observed low-frequency breathing dynamics.
 
\subsubsection{Lift-up Mechansim}

Another explanation for the amplification of the observed large-scale streamwise-elongated coherent structures is the lift-up mechanism. The lift-up effect describes how cross-stream disturbances amplify streamwise velocity perturbations \citep{Brandt2014TheFlows, SchmidHenningson2001}. In simple terms, streamwise vorticity mixes regions of slow and fast moving fluid, generating fluctuations in the streamwise velocity component. Typically, lift-up belongs to the class of non-modal mechanisms. Since the low-frequency breathing is associated with a stationary eigenmode, a purely non-modal origin appears unlikely. Nevertheless, the lift-up effect may still contribute to the amplification of large-scale structures and is therefore briefly discussed.
\cite{Marquet2009DirectNon} performed a global stability analysis on a DNS dataset of a laminar separation bubble behind a smooth backward-facing ramp. They identified a stationary, unstable three-dimensional mode where streamwise vortices generate alternating high- and low-speed streaks, characteristic of the lift-up mechanism.
Another relevant study by \cite{Pickering_LiftUp2020} examines lift-up, Kelvin–Helmholtz, and Orr mechanisms in turbulent jets using resolvent analysis and SPOD. Despite the different flow configuration, their visualization of the lift-up mechanism serves as a useful reference. Their results reveal streamwise-elongated structures in the low-frequency range when analysing the streamwise component of response modes. Cross-sectional visualizations display the streamwise velocity component along with cross- and spanwise velocity vectors, highlighting the underlying kinematics.

\begin{figure}
    \centering 
    \includegraphics[width=1.0\textwidth]{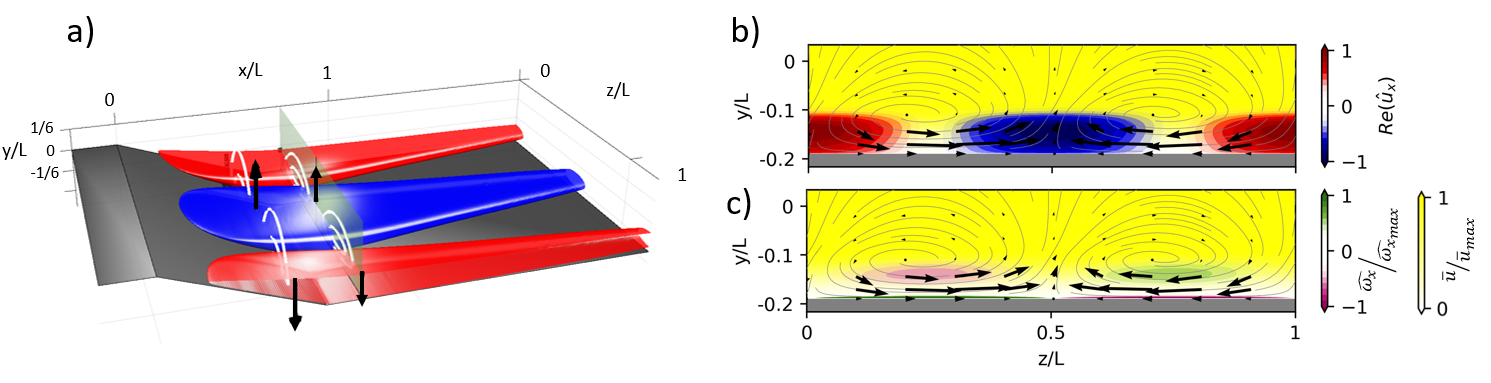}
    \caption{Leading resolvent mode for $St=0.01$ and $\beta L_b / 2\pi=1/3$ (or $L/\lambda_z=1$). Panel (a) presents 3D isosurfaces of the real part of the streamwise velocity ($u$) response mode, where blue and red indicate negative and positive values, respectively. Black lines represent streamlines of the cross-stream velocity components ($v$ and $w$) in a cross-section, with arrows denoting the direction of the $v$–$w$ response at selected locations. The transparent plane marks the reference cross-section visualised in panels (b) and (c). Panel (b) shows the streamwise velocity ($u$) response field in the cross-section, displayed as a blue-to-red contour plot. Arrows and streamlines illustrate the cross-stream ($v$–$w$) response field in the same plane. The background contour represents the normalised streamwise mean velocity $\overline{u}/\overline{u}_{max}$. Panel (c) depicts the normalised streamwise vorticity $\hat{\omega}_x/\hat{\omega}_{x, max}$, where $\hat{\omega}_x = (\partial\hat{w}/\partial y - \partial\hat{v}/\partial z)$, shown as a pink-to-green contour. Coordinates are scaled with the spanwise domain length $L$.}
    \label{fig:lift-up}
\end{figure}

Inspired by the work of \cite{Pickering_LiftUp2020} we analyze the leading resolvent modes with respect to the kinematic relationship between streamwise and crosswise velocity components.
Figure~\ref{fig:lift-up} shows the response mode for $St=0.01$ and $\beta L / 2\pi=1$. Panel (a) visualises the streamwise velocity component of the response mode as isocontours, revealing parallel streamwise-elongated structures, often referred to as streaks. Panel (b) shows a cross-sectional slice of the same response mode, corresponding to the plane indicated in the 3D visualization of panel (a). In this cross-section, in addition to the streamwise component, the cross-stream and spanwise velocity components ($v$ and $w$), as well as the streamwise mean flow velocity, are displayed.  
It is evident that the response mode is located within the shear layer formed between the low streamwise velocity region near the wall and the high-speed region in the free stream. The mode exhibits a clear kinematic pattern: regions of negative streamwise response coincide with upward motion away from the wall, transporting low-speed fluid toward the free stream. Conversely, high-speed streaks are associated with downward motion towards the wall. This alternating upward and downward motion generates a system of counter-rotating vortices. As shown in Figure~\ref{fig:lift-up}(c), the centers of these vortices, indicated by the maximum streamwise vorticity, are located between the streaks.
These observations suggest that the lift-up effect provides a plausible explanation for the formation of the observed streaks.

\subsubsection{Connecting the physical mechanisms}

The resolvent and LSA-based model successfully captures the key characteristics of the observed low-frequency dynamics, allowing for an investigation of the underlying physical mechanisms. The comparison with Görtler instability suggests that the characteristic spanwise wavelengths in the present setup are significantly larger than those predicted for Görtler vortices, making it unlikely that Görtler instability is the dominant mechanism. Instead, the presence of a stationary eigenmode at zero frequency, and the distribution of negative regions in the Rayleigh discriminant indicate that a centrifugal mechanism governs the low-frequency dynamics.

\begin{figure}
\centering
\includegraphics[width=0.6\textwidth]{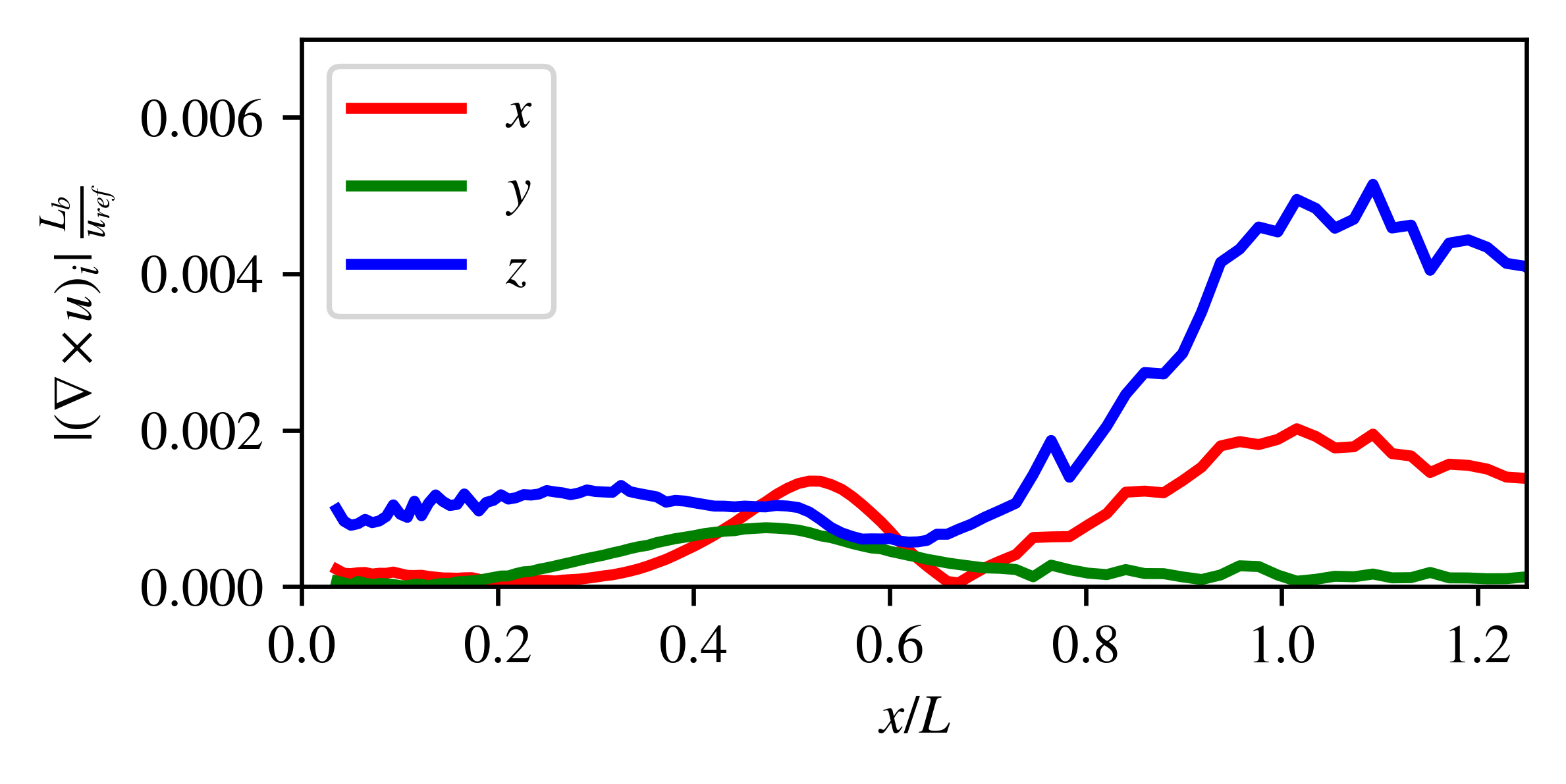}
\caption{Vorticity components of the leading resolvent mode for $St=0$ and $L/\lambda_z=1$ along the streamline shown in the upper panel of Fig.~\ref{fig:Rayleigh_Criterion}. The vorticity is non-dimensionalised using $\frac{L_b}{u_\infty}$. The axial coordinate is scaled with the spanwise domain length $L$.}
\label{fig:RA_vorticity_line_plot}
\end{figure}

To further examine the driving mechanism, Fig.~\ref{fig:RA_vorticity_line_plot} presents the vorticity components along the streamline indicated in Fig.~\ref{fig:Rayleigh_Criterion}. The profiles reveal a peak in the streamwise vorticity near the reattachment point, close to the region where the Rayleigh discriminant becomes negative and centrifugal effects are at play. Further downstream, the spanwise vorticity increases with the streamwise coordinate, a behaviour characteristic of streaks sustained by the lift-up mechanism. Similar observations were reported by \citet{SavarinoJFM2025} for a low-wavenumber case exhibiting signatures of both centrifugal and lift-up processes. This comparison suggests that the streaks are initiated by centrifugal effects and maintained by momentum transport across the reattached boundary layer, typical of the lift-up mechanism.

\section{Conclusion}

This study investigates the mode selection mechanisms in a TSB generated on a backward-facing ramp. In particular, we examine how periodic versus solid-wall boundary conditions affect the low-frequency dynamics of the flow. This leads to the related question of how the spanwise domain length influences the selection and structure of the dominant modes.

To address these questions, we conduct a global linear stability analysis, which reveals a zero-frequency eigenmode whose growth rate depends on the spanwise wavenumber. We then apply resolvent analysis to characterise the non-modal amplification mechanisms and associated coherent structures. Subsequently, we develop a standing wave model by analysing how boundary conditions constrain the spanwise wavenumbers. Periodic sidewalls restrict the flow to integer wavenumbers, with the longest wavelength limited by the channel width. Drawing an analogy with acoustics and surface waves, we model solid sidewalls as reflective, resulting in the formation of standing waves through the reflection of resolvent-based traveling modes.
Implementing this concept with a quasi-3D framework required simplifying assumptions. In particular, slip-wall boundary conditions are applied at the sidewalls, an approach that may seem unintuitive given the physical presence of no-slip walls. However, the SPOD results reveal that the fluctuating component attains its maximum amplitude near the sidewalls, suggesting that the slip-wall assumption provides a reasonable approximation in this context.

The SPOD analysis of the experimental data reveals low-frequency breathing dynamics in the symmetry plane for frequencies below $St=0.05$, dominated by a single mode that spans the entire separation bubble and forms a streamwise-elongated structure. In the horizontal plane, three modes are dominant at very low frequencies, exhibiting standing wave behaviour with distinct spanwise distributions of nodes and antinodes.
These experimentally observed structures are well captured by the standing wave resolvent model. The model successfully reproduces the spatial structure of the leading and two subleading SPOD modes, exhibiting strong agreement in both the symmetry and horizontal measurement planes.

Beyond reproducing the spatial structure, the model also offers insight into the underlying mechanism. The combined results from the global stability and resolvent analyses suggest that the low-frequency dynamics are most likely governed by a centrifugal mechanism. This interpretation is supported by the presence of a stationary eigenmode at zero frequency, regions of negative Rayleigh discriminant near separation and reattachment and characteristic vorticity. These observations indicate that curvature-driven centrifugal effects could play a dominant role in the breathing motion. Downstream of the bubble, however, the kinematic relationship between the streamwise and crosswise velocity components, together with the formation of counter-rotating vortices and alternating high- and low-speed streaks, points to an additional lift-up process. Taken together, the results indicate a plausible scenario in which the centrifugal mechanism initiates the low-frequency motion, while the lift-up mechanism amplifies the resulting streamwise-elongated structures further downstream.

The present results suggest that the selection of the dominant spanwise wavenumber of the low frequency dynamics is primarily influenced by the compatibility with the reflective sidewalls, while the growth rate or resolvent gain plays a secondary role. From a phenomenological perspective, the coherent structures that predominate are those whose wavelengths correspond to the spanwise channel width, rather than the one that exhibits the most significant streamwise amplification. In the context of future studies, it would be worthwhile to conduct a modal analysis and a resolvent analysis on a three-dimensional baseflow, incorporating the sidewalls, in order to investigate the physics of wall reflections.

In summary, our results show that the three-dimensional breathing dynamics can be effectively modeled on a two-dimensional mean flow if one takes wave reflection of the sidewalls into consideration. The standing wave model captures the breathing dynamics well. The present results further suggest that particular care should be taken when comparing experimental and numerical results, since the spanwise boundary conditions can significantly influence the observed dynamics.
This implies that for the low frequency dynamics of the turbulent separation bubble, the spanwise domain width should be considered as a relevant length scale, in addition to the bubble length.  

Our findings may have implications for computational fluid dynamics (CFD) studies. Periodic boundary conditions commonly employed in CFD simulations inherently limit the spanwise wavenumbers that can develop and are insufficient to capture the observed standing-wave behaviour. However, modelling full-span geometries with no-slip sidewalls substantially increases computational cost. Therefore, we suggest maximizing the spanwise domain size of CFD studies whenever possible and we propose free-slip sidewalls as a potentially feasible alternative for future investigation.

\section*{Acknowledgments}
The authors gratefully acknowledge the computing time made available to them on the high-performance computers at the NHR center NHR@ZIB and NHR@Göttingen. These centers are supported by the Federal Ministry of Education and Research and the state governments participating in the NHR (www.nhr-verein.de).

\section*{Funding.}
Funded by the Deutsche Forschungsgemeinschaft (DFG, German Research
Foundation) – project numbers 504349109, 506170981

\section*{Declaration of Interests.}
The authors report no conflict of interest.

\section*{Data availability statement.}
The data that support the findings of this study are available from the corresponding author, LMF, upon reasonable request. The code used for the linearized analyses is an open source software package that is available at \url{https://gitlab.com/felics-group/FELiCS}.

\section*{Use of artificial intelligence (AI) tools.}
Artificial intelligence tools (DeepL, ChatGPT, Perplexity) were used solely for language refinement.

\bibliographystyle{jfm}
\bibliography{references}

\end{document}